\documentclass{article}
\usepackage{graphicx}                    
\usepackage{color}                       
\usepackage{breakurl}                         
\usepackage{fancyhdr,natbib}
\usepackage{authblk}
\usepackage[a4paper]{geometry}
   
\usepackage{multirow}
\usepackage{longtable}
\usepackage{enumitem}
\usepackage{rotating}

\fancypagestyle{plainlower}{
  \fancyhf{}
  \fancyfoot[C]{\raisebox{-6\baselineskip}{\thepage}}
}

\begin{document}



\title{\bf JPEG2000 Image Compression on Solar EUV Images}


 
\author[1]{C.~E.~Fischer}
\affil[1] {Kiepenheuer Institut f\"{u}r Sonnenphysik \protect\\ Sch\"{o}neckstrasse 6, 79104 Freiburg, Germany}

\author[2] {D.~M\"{u}ller}
\affil[2] {European Space Agency/ESTEC \protect\\ Keplerlaan 1, 2200 AG Noordwijk, The Netherlands}

\author[3] {I.~De Moortel}
\affil[3] {School of Mathematics and Statistics \protect\\ University of St Andrews, St Andrews, Scotland, KY16 9SS, UK}
\date{}
\maketitle
 

\lhead{Fischer {\em et al.}}
\rhead{JPEG2000 Compression of Solar EUV Images}


%
\begin{abstract}
For future solar missions as well as ground-based telescopes, efficient ways to return and process data have become increasingly important. {\em Solar Orbiter}, {\em e.g.}, which is the next ESA/NASA mission to explore the Sun and the heliosphere, is a deep-space mission, which implies a limited telemetry rate that makes efficient onboard data compression a necessity to achieve the mission science goals. Missions like the {\em Solar Dynamics Observatory} (SDO) and future ground-based telescopes such as the {\em Daniel K. Inouye Solar Telescope}, on the other hand, face the challenge of making petabyte-sized solar data archives accessible to the solar community.
New image compression standards address these challenges by implementing efficient and flexible compression algorithms that can be tailored to user requirements. We analyse solar images from the {\em Atmospheric Imaging Assembly} (AIA) instrument onboard SDO to study the effect of lossy JPEG2000 (from the Joint Photographic Experts Group 2000) image compression at different bit rates. To assess the quality of compressed images, we use the mean structural similarity (MSSIM) index as well as the widely used peak signal-to-noise ratio (PSNR) as metrics and compare the two in the context of solar EUV images. In addition, we perform tests to validate the scientific use of the lossily compressed images by analysing examples of an on-disk and off-limb coronal-loop oscillation time-series observed by AIA/SDO.

\end{abstract}




                    


\section{Introduction}     \label{sec:introduction}

The data downlink rate of {\em Solar Orbiter} will be highly variable over time, scaling roughly with $1/r^2$, where $r$ is the spacecraft distance to Earth. Averaged over a 168-day orbit, it will return about 500\,MB of data per day, shared among 10 instruments. This is roughly $2.5$ times the data rate of the {\em Solar and Heliospheric Observatory} (SOHO) mission, but it pales in comparison to the 1.4\,TB per day of the SDO. Given this constrained telemetry return, implementing effective compression schemes is a necessity to achieve the mission science goals. The {\em Solar Orbiter Extreme Ultraviolet Imaging telescopes} (EUI), for example, will use an onboard compression algorithm to achieve this.  Current and past solar missions such as the {\em Transition Region and Coronal Explorer} (TRACE), SOHO, SDO and {\em Hinode} have also been using lossy JPEG compression at high-quality rates, as well as other compression schemes, to reduce telemetry volumes. Moreover, future ground-based instruments such as the {\em Visible Broadband Imager} (VBI) at the {\em Daniel K. Inouye Solar Telescope} (DKIST) will generate, even after data reduction and calibration, 350\,GB of data per day. In light of these numbers, lossy compression algorithms will also be key in the ``big data" regime to provide an efficient way to distribute and to browse these enormous data sets and to perform scientific data mining.

This approach is being used by the JHelioviewer \citep{MullerFDCAOWAHI09} tool, {\em e.g.}, which enables users to visually browse petabyte-scale data sets and makes use of the region-of-interest-based data access and decompression of JPEG2000 (from the Joint Photographic Experts Group 2000) encoded data. For AIA/SDO data, the lossy compression is performed at a bitrate of 0.5 bpp, which is entirely sufficient for visual data browsing, permits generation of running-difference movies and is even sufficient for certain types of scientific analysis.

Image processing using methods that combine lossless and lossy compression has, {\em e.g.}, been studied by \cite{2014arXiv1401.7433P}, who investigated the effects of lossy JPEG2000 compression on astronomical radio imaging data, and \cite{2005SoPh..228..253N} who tested image compression for solar EUV images. \cite{2005SoPh..228..253N} proposed a lossy preprocessing of the images by remapping the images into a lower bit depth, which reduces the precision in the image values but not beyond the calculated quantum noise, and  thereby, following a then lossless compression, achieves an overall higher compression rate. More recently, \cite{2016A&A...587A...9L} have investigated the effects of data compression when retrieving velocities with local correlation tracking from solar images. 

We aim to contribute to these investigations regarding the possibilities in solar image compression by investigating lossy solar image compression using JPEG2000 and by finding a suitable metric to quantify the quality of the compressed images and determine the implication for the scientific analysis of the compressed images.

In Section~\ref{sec:dataana} we introduce the JPEG2000 scheme with its advantages (Section~\ref{jp}), the images in our database we perform tests on (Section~\ref{datab}), and we elaborate on the quality metrics selected for comparison in Section~\ref{quame}. Section~\ref{compeff} is devoted to the study of the effect of compression on the image resolution and the artefacts introduced by the compression. We then compare two quality metrics and their prediction of image quality, specifically for solar EUV images and the structures seen in them in Section~\ref{mssim_psnr}. Section~\ref{sec:resultsclo} gives a first look at actual science cases and the change in results caused by compression errors. For this we choose two coronal oscillation events observed at the limb (Section~\ref{sec:resultsclo_1}) and on the disk (Section~\ref{sec:resultsclo_2}). In Section~\ref{sec:concl} we summarise our findings. 
                    


\section{Compression Scheme, Test Database, and Quality Metric Definition}    \label{sec:dataana} 
\subsection{JPEG2000} \label{jp}

The JPEG  was the first standardised compression algorithm. The later developed JPEG2000 is defined in the ISO/IEC 15444-1:2004 document and is described, for example, in \cite{Skodras01thejpeg}. It provides lossy and lossless compression, and several new features were introduced, such as progressive decoding, which allows for more customised extraction. We chose the JPEG2000 compression scheme as it is an ISO standard, highly adaptable to user needs, and is also implemented in the Interactive Data Language\textsuperscript{\textregistered}\footnote{ www.exelisvis.com} that is widely used in astronomy.\\

The algorithm comprises the following steps:
\begin{enumerate}[label={(\arabic*)},itemsep=0.0cm]
\item Preprocessing: Images are optionally tiled to be processed individually, unsigned image values are shifted to be symmetric around 0,  and an optional colour transformation from RGB (red, green, blue) to YCbCr is performed.
\item DWT: A discrete wavelet transform is applied by recursively passing high- and low-frequency filters. The tiles are decomposed dyadicly resulting in sub-bands for each tile. For lossless compression, a reversible wavelet transform is used, while for lossy compression,  the wavelet transform is irreversible.
\item Quantization: At this stage, precision in the wavelet coefficients is reduced if lossy compression has been chosen. 
\item Bit encoding: The data are stored in progressively higher precision in so-called bitplanes. 
\end{enumerate}

In IDL, an IDLffJPEG2000 object has been implemented. This object class accepts keywords specifying the number of tiles, bit rates, and other parameters of the compression algorithm. It uses the Kakadu\footnote{ www.kakadusoftware.com} code to compress the images. In addition, open source implementations like OpenJPEG\footnote{http://www.openjpeg.org/} exist.

\subsection{Database} \label{datab}
 We assembled a database by selecting images at arbitrary dates from AIA/SDO in several wavelengths and choosing quiet-Sun and active-region targets with exposure times of $2$\,s. 
  For details on this instrument see~\cite{2012SoPh..275...17L}. Table~\ref{database} in the Appendix lists the various datasets taken in rapid succession within seconds at different wavelengths and their properties, such as the observing time. The Level 1.5 AIA/SDO images were obtained using SolarSoft\footnote{www.lmsal.com$\slash$solarsoft} commands provided by the instrument team of the Lockheed-Martin Solar and Astrophysics Laboratory (LMSAL). These data are dark- and flat-fielded, passed through a bad-pixel-removal algorithm, are co-aligned, rotation-corrected, and stored as 16-bit integer FITS files. We do not cover the details of the preprocessing of the images here, although this has to be taken carefully into account when designing the onboard compression on a satellite such as {\em Solar Orbiter}, to avoid encoding unnecessary data.

Compression is performed on each image individually. We choose one level (set by the {\bf \em n\textunderscore level} keyword) and one layer (keyword {\bf\em bit\textunderscore rate}) in the IDL JPEG2000 object class. The {\bf\em bit\textunderscore depth} keyword is set to 16. We note that this is not the bit rate in the compressed image. We obtain in$jp2$ images with selected bit rates in this way.

\subsection{Quality Metrics} \label{quame}

According to~\cite{Wang_MSE}, one of the main reasons why the mean square error (MSE) and the peak signal-to- noise ratio (PSNR) derived from the MSE are the most commonly used quality metrics is their simplicity and convenience. More recently developed quality metrics, such as the structural similarity index (SSIM)  and its derivatives, have the advantage (compared to the MSE) of being optimised for human eye perception and, as the name indicates, take the interdependency of close by pixels into account  in creating perceived structures. We choose the SSIM as a quality metric to contrast to the MSE because a large part of solar research still relies on visual inspection of solar data and manual event selection as a first step. The SSIM has previously been proposed as a quality measure by \cite{Gissot:2907}, who suggest that it may outperform the traditional MSE metric in assessing the image quality of solar EUV images. Especially for the studied EUV images with easily recognisable loop structures, this metric seems more appropriate. Additional deciding factors were the short computation time, relative simple algorithm, and the proven applicability of the SSIM and its derivatives to a wide range of topics in image processing.

\subsubsection{MSE and PSNR}
The MSE is a measure of the mean difference between the image pixel values between two images. With $u$ the original uncompressed image and $v$ being the compressed image, with both size ($M,N$) and coordinates $x$ and $y$, the $MSE$ is defined as 

\begin {equation}
{MSE}=\frac{1}{M N}\sum_{x=1}^{M}\sum_{y=1}^{N} \left( u_{xy}-v_{xy} \right) ^{2}.
\label{mse_eq}
\end{equation}

The PSNR takes the dynamic range $L$ of all the pixel values in the image into account and is derived from the $MSE$ as
\begin {equation}
{PSNR}=10\,  {log} \left(\frac{L^{2}}{MSE} \right) .
\label{psnr_eq}
\end{equation}

\subsubsection{SSIM and MSSIM}

The general algorithm for the SSIM index was defined, tested, and validated by \cite{Wang04imagequality} using a database of JPEG2000 compressed images and the Mean Opinion Scores (MOS) of human test subjects.
It defines three independent image characteristics: luminance, contrast, and structure.
The luminance measures the likeliness of the mean value between two images, whereas the contrast compares the standard deviations. The structure term is determined by the correlation between two images $u$ and $v$ and measures the tendency of $u$ and $v$ to ``vary together, [and is] thus an indication of structural similarity" \citep{multiscalewang}. The overall similarity measure is a combination of these three image characteristics weighted with a weighting function $f$:

\begin{equation}
SSIM_{({u},{v})}=f(l_{({u},{v})}c_{({u},{v})}s_{({u},{v})}).
\label{SSIMeq_global}
\end{equation}

We follow the approach of \cite{Wang04imagequality} here and weight the three parts of the SSIM (luminance, $l$, contrast, $c$, and structure comparison, $s$) equally. For images $u$ and $v$, 
with $\sigma$ being the standard deviation and $\mu$ the mean intensity, they are defined as follows:

\begin{equation}
l_{({u},{v})}=\frac{2\mu_{u}\mu_{v}+C_{1}}{\mu_{u}^2+\mu_{v}^2+C_{1}},
\label{luminance}
\end{equation}

\begin{equation}
c_{({u},{v})}=\frac{2\sigma_{u}\sigma_{v}+C_{2}}{\sigma_{u}^2+\sigma_{v}^2+C_{2}},
\label{contrast}
\end{equation}

\begin{equation}
s_{({u},{v})}=\frac{2\sigma_{uv}+\frac{C_{2}}{2}}{\sigma_{u}\sigma_{v}+\frac{C_{2}}{2}}.
\label{structure_comparison}
\end{equation}

The constants $C_{1}=K_{1}L^2$ and $C_{2}=K_{2}L^2$ have low values and are included to avoid instability caused by division through zero. 
 The  low values $K_{1}$ $\ll 1$ and $K_{2}$ $\ll$ 1 also take the dynamic range $L$ of the image into account. \cite{Wang04imagequality} empirically determined the values for $K_{1}$ and $K_{2}$ to be $0.01$ and $0.03$, respectively. The authors found these values by using $8$-bit grayscale images with a dynamic range of 255. In our case, we have a dynamic range of $L=65535$ with $16$-bit images. As the maximum in the images is usually around $10^{4}$, we had to adjust the K factors by $\left ( L_{8bit}/L_{16bit} \right)^{2}$ to ensure that these factors do not dominate the nominator.

To account for the locally varying image structure, \cite{Wang04imagequality} introduced a windowing algorithm. Instead of defining the SSIM globally, a sliding Gaussian-weighted window, with 11 x 11 pixel, for example, is moved over the image pixel by pixel. The SSIM is calculated for the image patches $u_{j}$ and $v_{j}$  of the $j$-th window:

\begin{equation}
SSIM_{{u}_{j},{v}_{j}}=l_{({u}_{j},{v}_{j})}c_{({u}_{j},{v}_{j})}s_{({u}_{j},{v}_{j})}.
\label{SSIMeq}
\end{equation}

By performing for each pixel, we produce a 2-D SSIM. Finally, the mean over the whole SSIM map is the mean $SSIM$ (MSSIM):

\begin{equation}
MSSIM_{({u},{v})}=\frac{1}{M}\sum_{j=1}^{M}SSIM_{{u}_{j},{v}_{j}},
\label{MSSIM}
\end{equation}
where $M$ is the number of local windows. This is then again a single-value quality parameter.

                    


\section{Results}\label{sec:results}

\subsection{Compression Effects}
\label{compeff}
We study the effect of image compression by comparing the intensity values, the intensity distribution, and the frequency content of the images before and after compression. We are especially interested in any blurring effects that diminish the visible structures such as loops or mossy areas. While the loops are easily identified as clear strands, moss has in contrast a ``spongy" appearance and is thought to be the upper transition region emission of hot coronal loops with the presence of chromospheric jets or spicules interspersing these EUV emission elements~\citep{1999SoPh..190..409B}.

\begin{table}
\caption{The table lists the chosen wavelength and region in the first column, followed by the mean intensity in the entire image in data units. The images were compressed with the JPEG2000 scheme. In the remaining columns we list the relative compression ratio defined as the ratio between the bitrates for lossless and lossy compression, the bitrate (bits per pixel), the MSSIM value, and finally the  achieved PSNR. }
\label{imdegtable}

\begin{tabular}{l c c c   c c}
 data & mean &CR$_{rel}$ & bits per &  MSSIM & PSNR   \\ 
& intensity [DN] & & pixel &  & \\ 
 \hline
 \hline
 AIA 171 &982.88&   1 (lossless)  &      7.64    & &   \\ 
AR    &   & 3  &      2.55    &    0.95 &   77.09   \\ 
     &   &    15    &  0.51   &   0.75   &    65.92 \\ 
 &   &    25     & 0.31     &     0.68   &    64.04
\\ 
  \hline
AIA 171&  233.06  &   1  (lossless)   &  6.38 & & \\ 
 QS  &  &    3    &   2.13    & 0.95  &   80.95   \\ 
&   &  15   &   0.43   &   0.72   &  73.83\\ 
&    &  25   &   0.26     &  0.62   &  72.25 \\
 \hline
   AIA 304 &  113.42  &    1 (lossless) & 5.62 & & \\ 
AR   &   &     3   &    1.87      &   0.88   &  85.19 \\ 
&      &    15   &     0.37   &   0.58  &   76.04 \\ 
   &     &    25  &      0.22     & 0.49   &  74.13  \\
   \hline
  AIA 304  &   51.33    &  1 (lossless) & 4.93  & & \\ 
 QS   &   &    3     &  1.64      &  0.90&     87.18 \\ 
  &     &    15     &   0.33        & 0.59 &    80.91   \\ 
   &     &        25     &     0.20     & 0.48  &   79.56 \\
  \hline
\end{tabular}
\end{table}

 \begin{figure*}
\includegraphics[width=\textwidth]{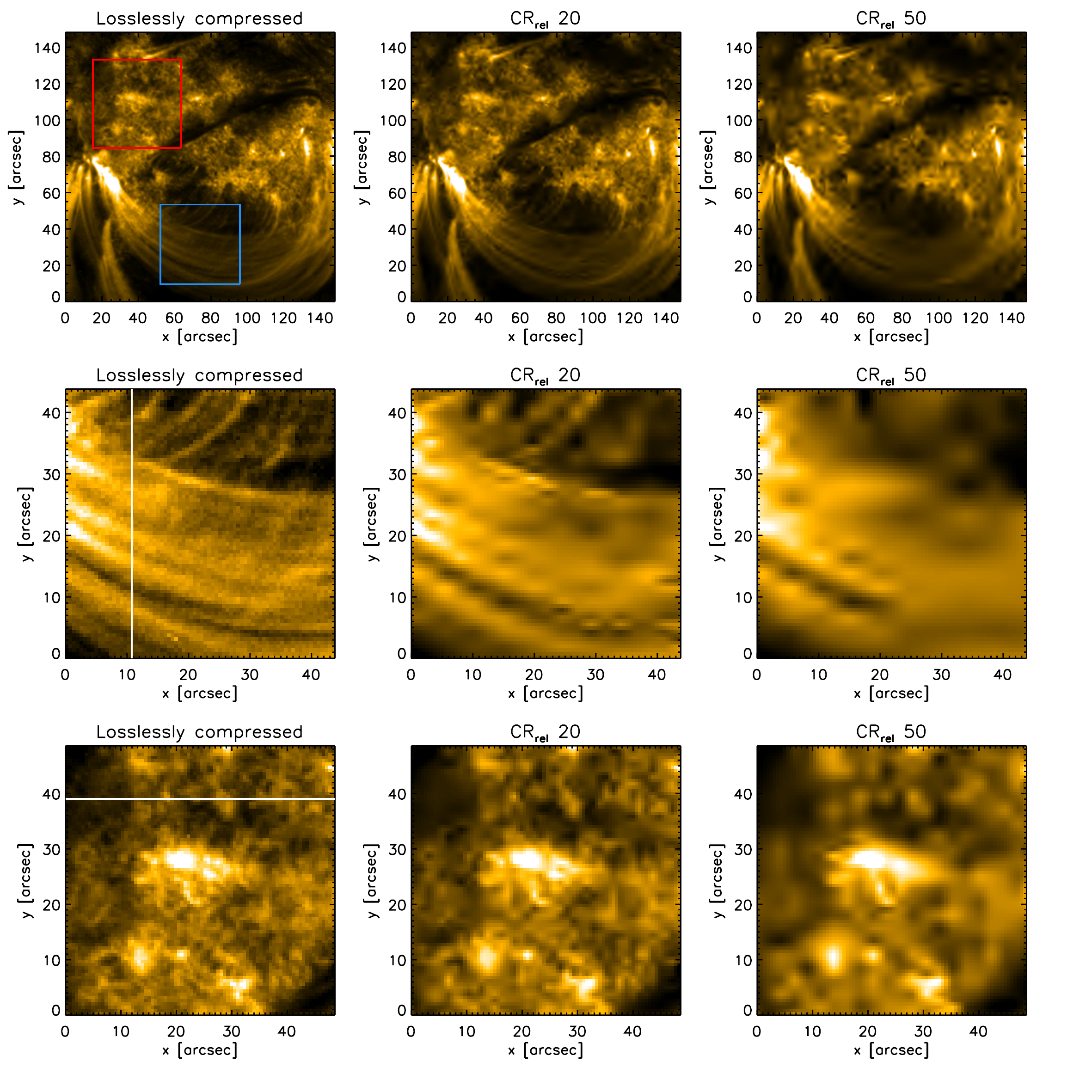} 

  \caption{The rows display extracted subimages from an AIA 171\,\AA\, full-disk image from 4 January 2012 with increasing compression (second column at CR$_{rel}$ 20, third column at CR$_{rel}$ 50). The blue box in the upper image marks the region displayed in the second row, and the red box the region in the last row. The white lines marked in the left images of the second and third row indicate cross sections that are analysed further in Figure~\ref{imgcont2}. 
}
 \label{imgcont1}
\end{figure*}


 \begin{figure*}
\includegraphics[width=0.8\textwidth]{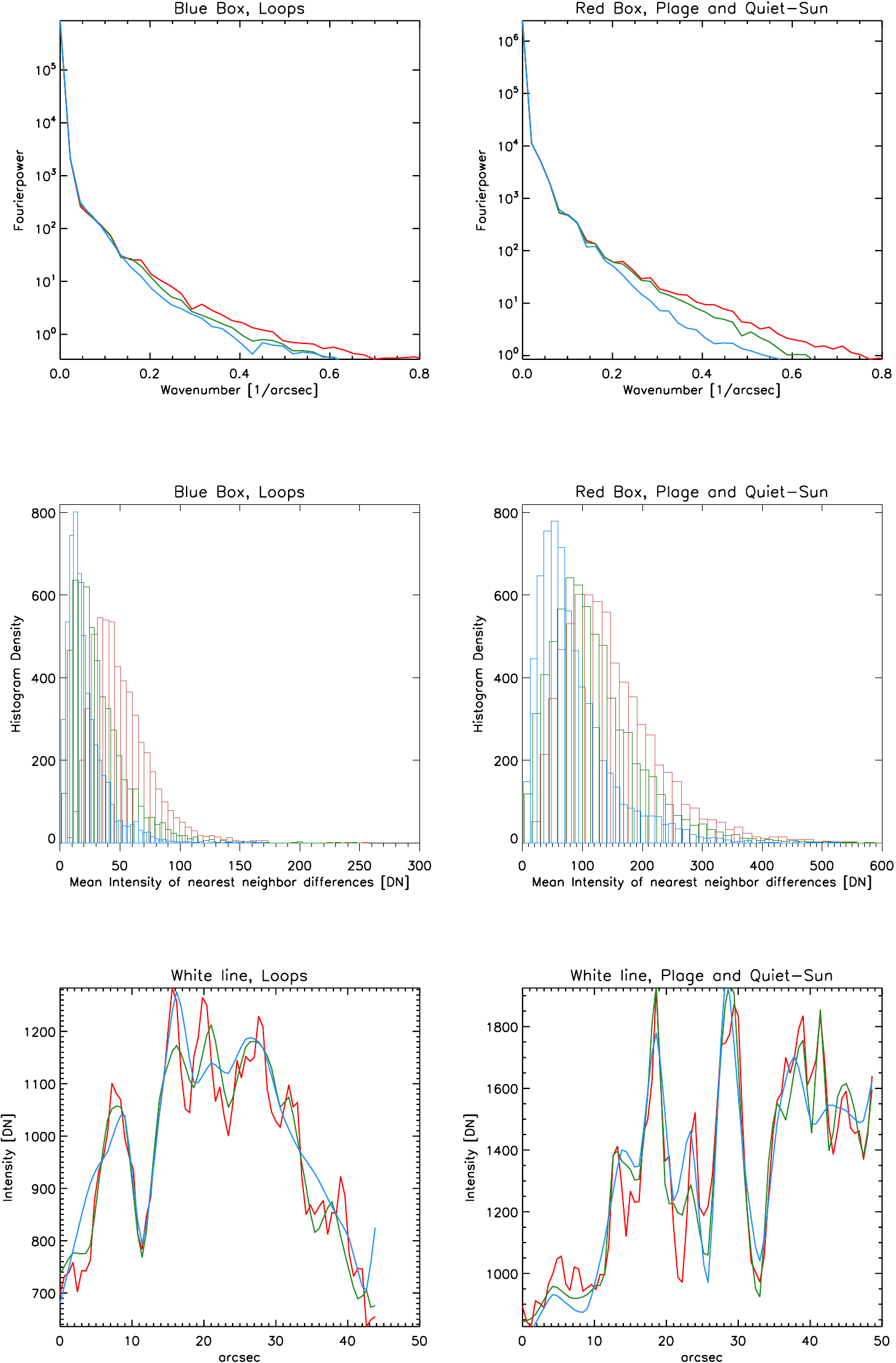} 
  \caption{ Left column: Spatial power spectrum for the region in the blue box indicated in Figure~\ref{imgcont1} followed by the mean of the nearest-neighbour difference, and finally, intensity values of the cut marked with white lines in Figure~\ref{imgcont1}. The red lines are for the losslessly compressed image, green for the images compressed with a relative compression ratio of 20, and blue for a  relative compression ratio of 50. 
   Right column: Same as in the left column, but now for the plage - and quiet-Sun region marked with a red box in Figure~\ref{imgcont1}.  
}
 \label{imgcont2}
\end{figure*}


For equal dynamic range and spatial size of the image, images at 171\,\AA\,and 193\,\AA\,generally require more storage space when compressed using lossless JPEG 2000 than the images at 211\,\AA\,and 304\,\AA\,. The bit depth of 16 is only a nominal bit depth with the actual image intensity values not using the full available dynamic range. Images with a lower dynamic range will require less storage after lossless compression. This is also shown in Table~\ref{imdegtable} where we list the bits per pixel (bpp) for lossless and increasing compression for the dataset taken on 4 January 2012 of an active region and a more quiet-Sun region at 171\,\AA\, and 304\,\AA\,. 
Throughout this article, we therefore use the relative compression ratio (CR$_{rel}$), which is defined here by the file size of the losslessly compressed image (lossless compression using the {\bf\em reversible} keyword)  divided by the file size of the lossily compressed image. 
The bits per pixel (bpp) are obtained by dividing the file size by the number of pixels in the image. This allows us to compare between the different wavelengths.  We chose this definition of relative compression ratio because we are interested in the actual storage space that is saved by lossy compression.

In Figure~\ref{imgcont1} we have selected an active region displaying loops as well as plage and quiet-Sun areas.  In the first row we show the range from the losslessly compressed image to a relative compression ratio of 20 (second panel) up to a relative compression ratio of 50 (last panel). Even at such high compression the structures are recognisable, but at close inspection (second and third row), the compression effects are clearly visible. JPEG2000-compressed images do not suffer from the blockiness of JPEG compressed images because the transforms are usually applied to the entire image and not on tiles, as with the JPEG code. However, so-called ringing artefacts, resulting in less clearly defined edges at sharp transitions, and a general image blur become visible to the naked eye at high compression. In a region dominated by loops (outlined by the blue box in the upper left image and shown in the second row), an overall blur and additional structure with stripes almost perpendicular to the loops is visible.  In the more quiet area (outlined by the red box in the upper left image and shown in the third row), the small-scale intensity variations disappear, and only the very bright structures remain discernible.

In Figure~\ref{imgcont2} we confirm the visual result by analysing the boxed regions and cross sections of the boxed regions (cuts marked with white lines in the left images in the second and third row of Figure~\ref{imgcont1}). 
The top panels in Figure~\ref{imgcont2} show the magnitude of the spatial Fourier transform (spatial power spectrum) plotted over constant spatial wavenumber, $k$, with $k=k_{x}^{2}+k_{y}^{2}$, resulting in a 1D power spectrum. In both cases (the region displaying loops from Figure~\ref{imgcont1} and the plage- and quiet-Sun region from the same image), the higher frequencies disappear with increasing compression, implying the loss of fine-scale structure. For the plage- and quiet-Sun region, which also has a larger high-frequency content to begin with, this effect is stronger.
The mean intensity (value of the spatial power spectrum at 0 spatial frequency) does not vary for the relative compression ratios chosen. This is also reflected in the stability of the luminance term in the SSIM-map calculation.

 In the second row, we plot the histograms of the mean of the nearest-neighbour difference, which gives an indication of the intensity gradients in the image. The intensity differences decrease, resulting in a shift of the maximum in the histogram to lower values for both regions, indicating blurring of the image. In the loop region this occurs at a faster relative rate, and the histogram  also becomes narrower at high compression. This means that the intensity differences are less widely distributed, which most affects regions with lower intensity gradients and lower intensity values.

The last row displays cuts through the regions. For the loop regions we chose a cut more perpendicular to the loops to distinguish between individual loops. The loops take up a horizontal scale of about 5\,arcsec. Even with a relative compression ratio of 50, we can clearly still discern the individual loops, although their edges are less steep and intensity changes within a loop are diminished. In contrast, the structure in the plage and quiet region is contained on scales of around 2\,arcsec and is consequently blurred out by a compression at the same relative compression ratio.

It is therefore clear that when determining the image blurring by JPEG2000 compression, the relevant scales of the image structures and therefore the pixel scale of the spatial sampling, but also the intensity gradients, need to be taken into account. In the studied images with a scale of 0.6\,arcsec per pixel, the fine-structure content of plage- and quiet-Sun areas prohibits JPEG2000 compression with high relative compression ratios. Images showing EUV loops can tolerate a higher relative compression ratio when the only interest is identifying the loops. In Section~\ref{sec:resultsclo} the corresponding limits to the compression rates are analysed for studying loop oscillations in solar EUV images where the loop width and precise locations become important.

\subsection{Comparison of MSSIM and PSNR} \label{mssim_psnr}

 \begin{figure}
\includegraphics[width=\textwidth]{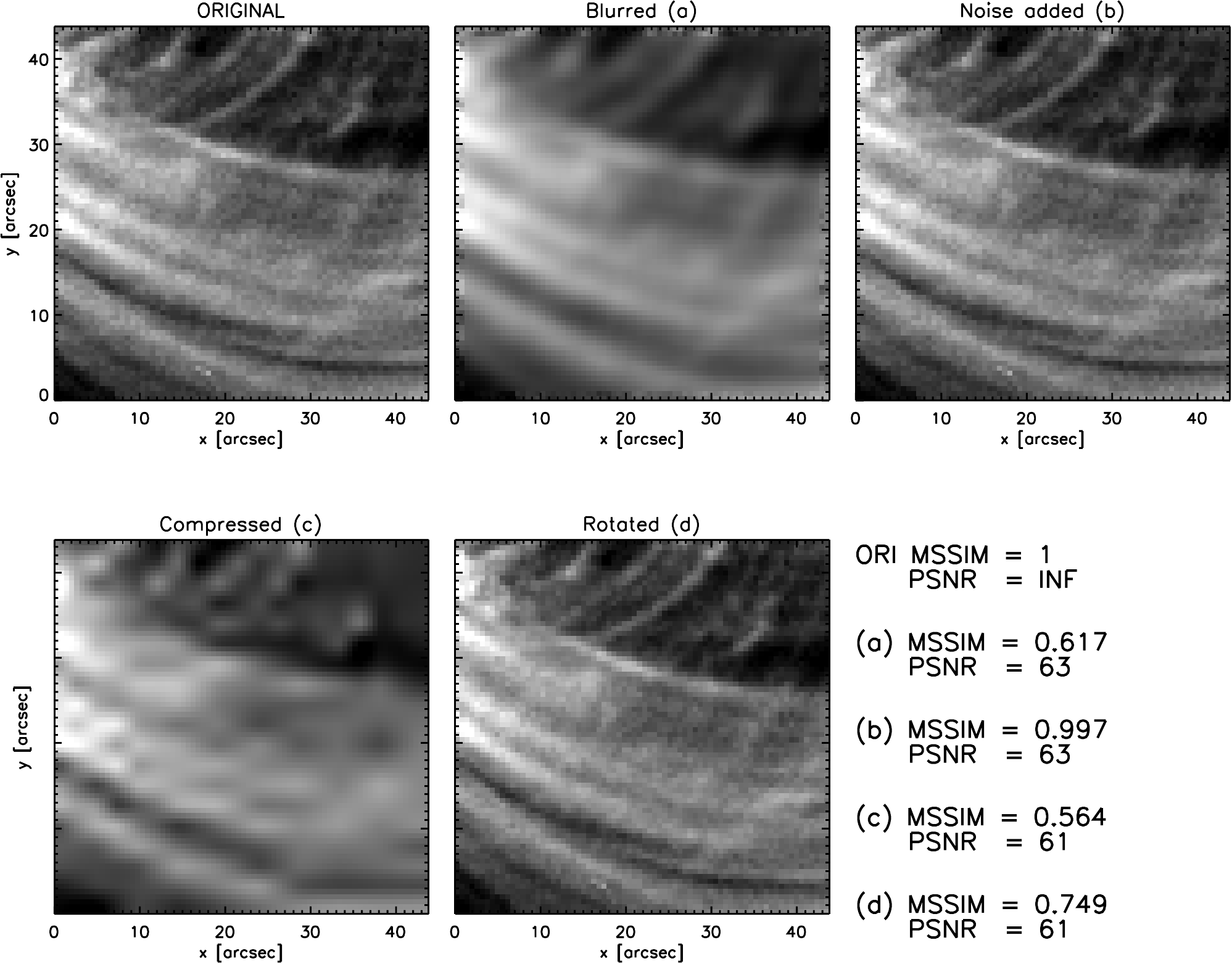}
  \caption{Top row, left to right: Original AIA/SDO 171\,\AA\, image, (a) blurred with the IDL function { \bf {smooth$.$pro}}  with keyword width $=4$ and (b) with 5$\%$ Poisson noise added using the IDL function {\bf {poidev$.$pro}}.
Bottom row, left to right:  Same image (c) JPEG2000 compressed with a CR$_{rel}$ of $20.5$ and  (d) rotated counterclockwise by 3 degrees using the IDL function {\bf {rot$.$pro}} with the keyword cubic $=-1$. On the right we list the corresponding MSSIM and PSNR values for the different image distortions.}
 \label{imdist}
\end{figure}

 \begin{figure*}
\includegraphics[width=\textwidth]{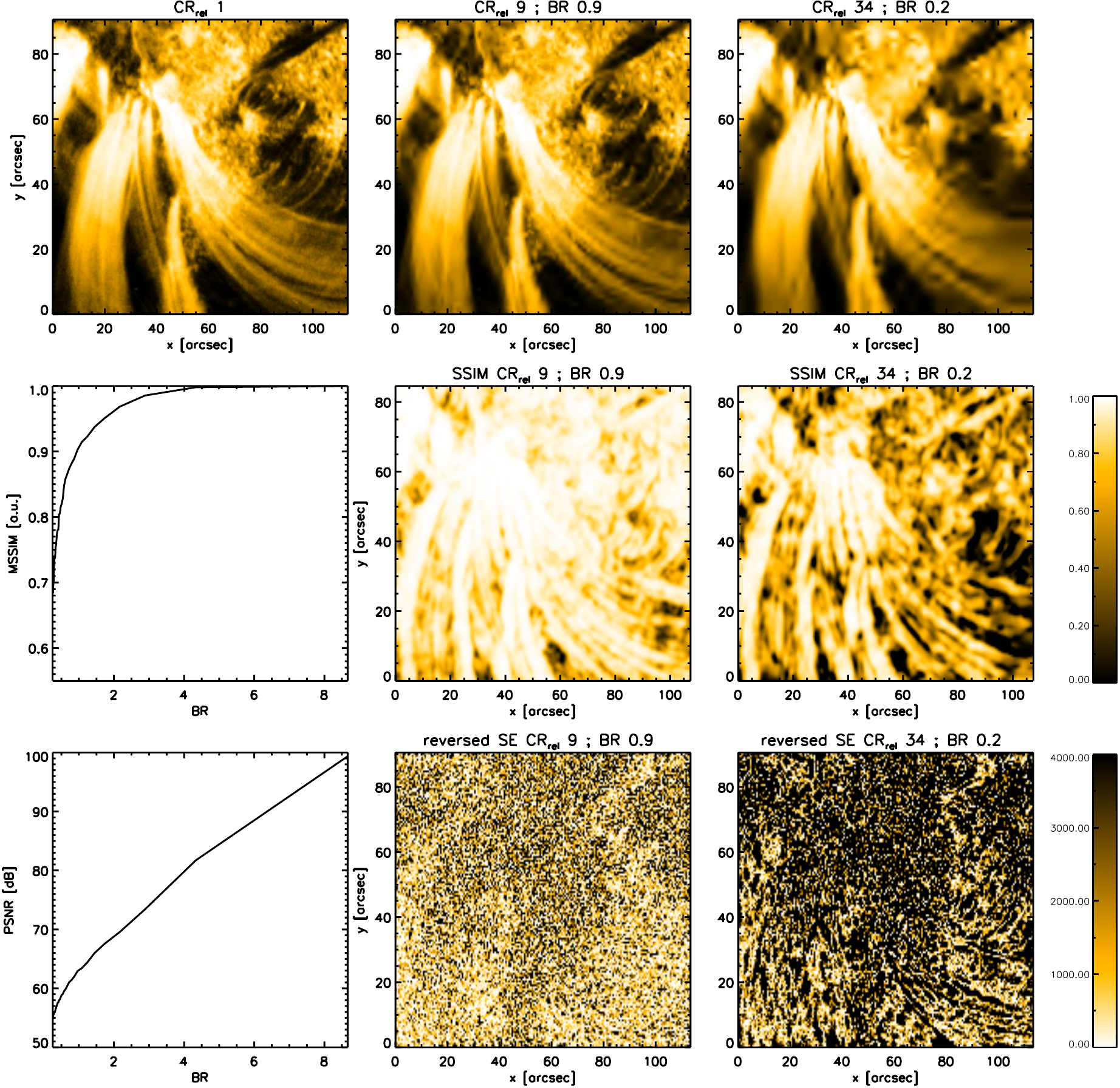} 
  \caption{First row: Uncompressed and compressed AIA/SDO 171\,\AA\,subimage for relative compression ratio (CR$_{rel}$) 1, 9 and 34 corresponding to bitrates (BR) $8.6, 0.9$ and $0.2$. 
  Second row, left to right:  MSSIM plotted against bitrate (BR) for the image in the first row.  SSIM maps for the two relative compression ratios scaled into the same range. Bright areas signify good correlation between the uncompressed and compressed image.
  Third row, left to right: PSNR plotted against bitrate (BR) for the image in the first row. Squared error maps for the two relative compression ratios scaled into the same range. The values have been multiplied by $-1$ with bright areas  again identifying a good correlation between the uncompressed and compressed image.}
 \label{maps}
\end{figure*}


A quality metric should not only confirm the subjective impression of the quality of the compressed image, but also reflect the artefacts introduced by the chosen specific compression scheme. One of our aims therefore was to compare the performance of the more widely used PSNR to the MSSIM index when measuring the degradation of structures such as loops in solar EUV images that are due to compression. 
The last two columns in Table~\ref{imdegtable} list  the MSSIM as examples for different regions and wavelengths, which a range of between 0 and 1 and the PSNR in decibels. The MSSIM and the PSNR cannot be directly compared because they operate on different scales. However, we can draw conclusions from their sensitivity to an increasing relative compression ratio. 

 In Figure~\ref{imdist}, we demonstrate the effect of blurring, adding noise, compression, and rotation on the PSNR and MSSIM metrics.
The PSNR is reduced to a similar value for all types of distortions, whereas the MSSIM is less sensitive to noise being added (panel b) and still finds the structure in the very slightly rotated image (panel d) whereas heavily compressed (panel a) and blurred images (panel c) result in low MSSIM values. The PSNR is more suitable for detecting additional noise, such as the compression quantisation noise (a side effect of the precision cut-off during compression) in an image and is not only sensitive when blurring occurs (the dominating side effects of the JPEG2000 compression when determining the visual image quality), in contrast to the MSSIM. 

In the first row of Figure~\ref{maps} we  first show a losslessly compressed region taken with AIA/SDO in the 171\,\AA\,wavelength that shows coronal loops, plage, and also more quiet regions. We then choose a relative compression ratio of 9 (second figure in the first row of Figure~\ref{maps}) and 34 (last figure in the first row of Figure~\ref{maps}), and blurring and the ringing artefacts that are due to increasing JPEG2000 compression are again clearly visible. 
The second and third rows of Figure~\ref{maps} show the MSSIM curve and the PSNR curve and their corresponding maps, the  2D SSIM map obtained by using Equation~(\ref{SSIMeq}) for the Gaussian windows, and the squared error (SE) map, which is the difference between the original image and the compressed images, for the two selected relative compression ratios. 

In contrast to most figures in this article, in this case we plot the MSSIM and PSNR curves {\em versus} bitrate instead of relative compression ratio, as the compression ratio (bitrate of losslessly compressed image to bitrate of compressed image) goes with the inverse of the bitrate and we are here interested in the direct response of the curve to decreasing bitrate. 
Starting at a high bitrate for the losslessly compressed image, the PSNR decreases linearly with bitrate. The MSSIM in contrast shows a plateau at high MSSIM values before decreasing rapidly. The turning point is around a bitrate of two.

The maps reveal that even for low relative compression ratios (first map), the SSIM indeed picks up on the structure in the image by giving areas with clear structures that can be visually recognised (loops) a high (good) quality value, whereas the SE maps show more of a random appearance affecting all pixels. The SE map and therefore the PSNR is highly sensitive to the quantisation noise of the JPEG compression, which is superimposed onto the image, but does not affect the structures such as loops as much as the blurring does. At high compression (CR$_{rel}$ $=34$), the SE maps also starts to trace the loops, and the patchy appearance in both the SSIM map and the SE map reveal the ringing artefacts.

\subsection{MSSIM for Different Wavelengths and Regions}\label{wvl_mssim}

\begin{figure}
\includegraphics[width=\textwidth]{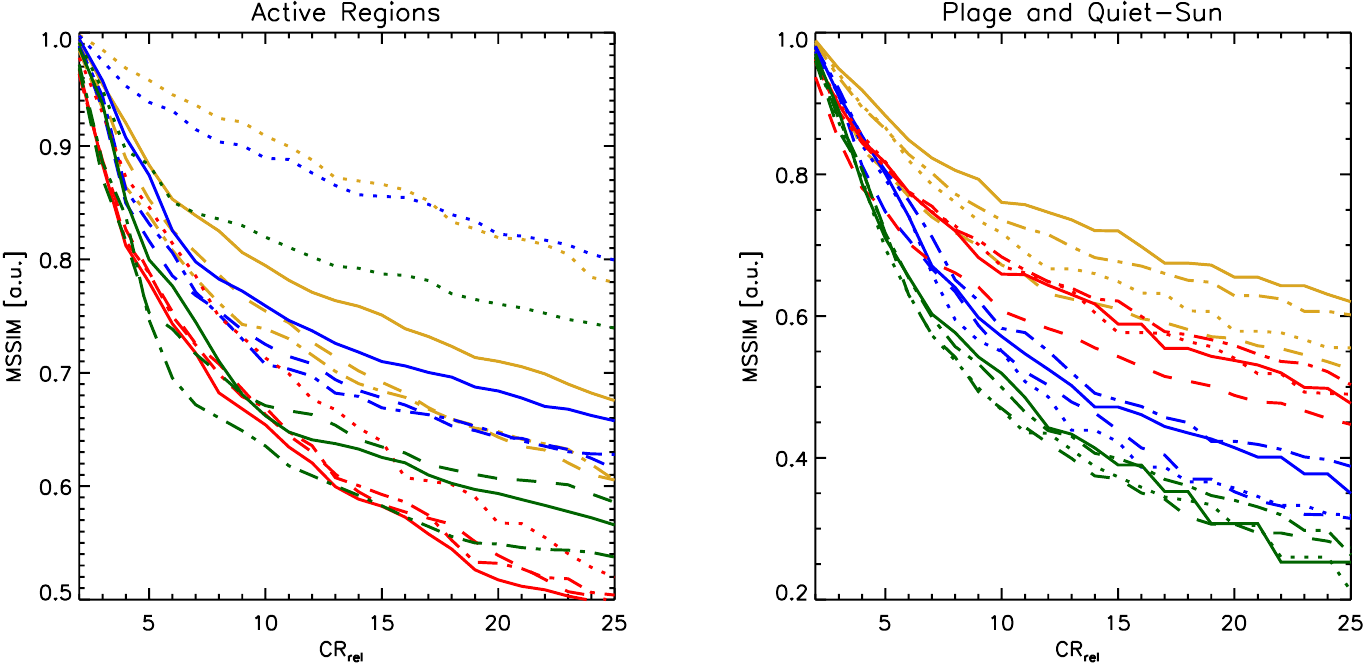} 

  \caption{MSSIM curves for AIA/SDO images in the wavelengths: 171\,\AA\, in yellow, 193\,\AA\, in blue, 304\,\AA\, in red, and 211\,\AA\, in green. The solid lines are results for images from the 4 January 2012 dataset, dotted for the 4 January 2013  dataset, dashed for the 19 January 2012  dataset, and dash-dotted for the  28 March 2013 dataset. We note that the scale range in the y-axis differs between the two panels.
}
 \label{diff_wvl_ssim}
\end{figure}

In Figure~\ref{diff_wvl_ssim} we plot the MSSIM value for increasing relative compression ratios for different wavelengths of the AIA/SDO images shown in Table~\ref{database}.
From the plot in the left panel it is clear that MSSIM curves between different active regions for the same wavelengths can greatly differ. However, for a single region, the MSSIM curves compared between wavelengths follow the same order. The 171\,\AA\, images, for example, show consistently higher MSSIM values. Several factors play a role when determining the MSSIM curve: the wavelength influences the size of the structures, the intensity values, and gradients, but the choice of region (for example, more loop-filled areas) is important as well. The region chosen for the dataset of 4 January 2013, for example, is entirely filled with coronal loops and can endure higher compression rates and maintain high MSSIM values for all wavelengths. This corresponds well with the findings in Section~\ref{compeff}, where by studying the results of image compression on different solar structures, we found that the loops were more compressible, which confirms that the MSSIM is a good quality metric.  
 The images in 211\,\AA\, and 304\,\AA\, exhibit similar MSSIM values for the active regions, but behave differently for the more quiet and plage regions (right panel in Figure~\ref{diff_wvl_ssim}). Again, this is the result of loops being more defined and structured in 211\,\AA\, relative to 304\,\AA\, for active regions, whereas the fine scale structure of plage and the quiet Sun is better resolved in 304\,\AA\, than in the low intensity noise-like features seen in 211\,\AA\, for the same region. The images in Table~\ref{database} in the Appendix show for confirmation the different regions in the various wavelengths as examples of the solar structure that is to be expected. 

\begin{figure}
\includegraphics[width=\textwidth]{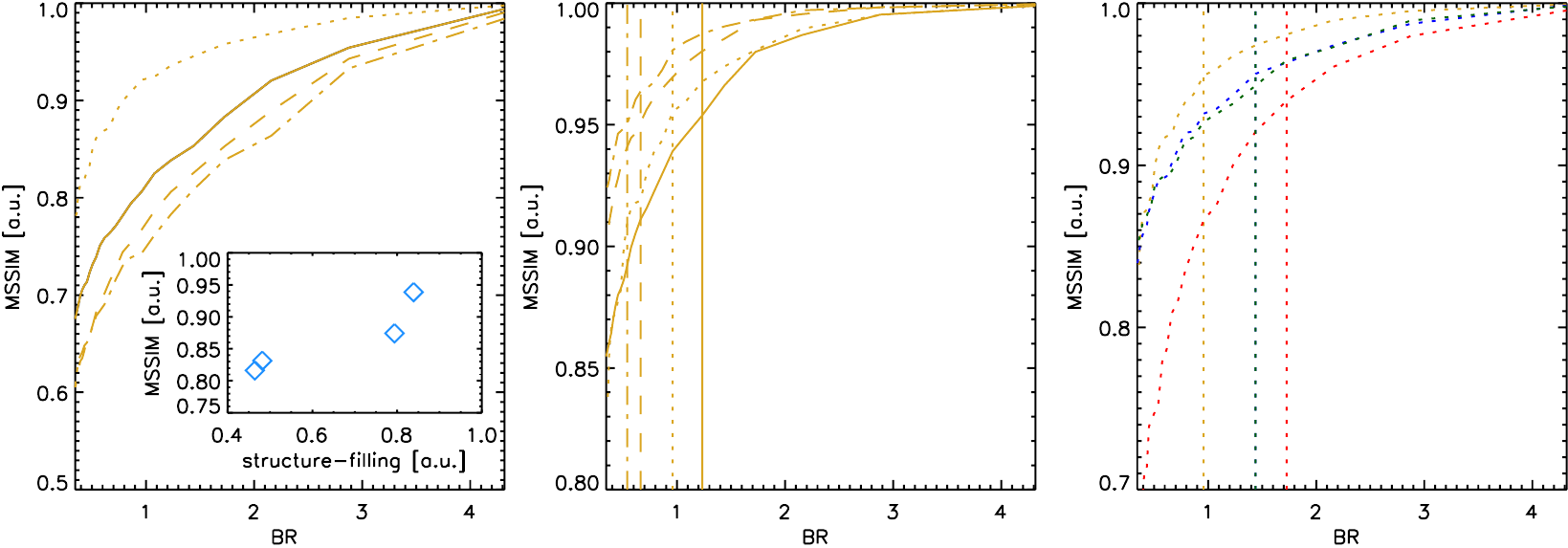} 

  \caption{MSSIM curves for AIA/SDO images. Left to right: In the first panel we only show the MSSIM for the four active regions in AIA 171\,\AA\,, now {\em versus} bitrate curves instead of the relative compression ratio in Figure~\ref{diff_wvl_ssim}. The different line styles correspond to the line styles chosen in Figure~\ref{diff_wvl_ssim} for the different dates. The inset shows for a CR$_{rel}$ of three (diamonds) the MSSIM {\em versus} a structure-filling parameter (see text for explanation). The following panel shows the MSSIM curves for the same regions, now only for the pixels within the mask created with a cut-off of 1000 DN. We note that the y--axis range has been changed compared to the first panel to display the curves more clearly. The vertical lines denote the bitrate at which a MSSIM of 0.95 is achieved. The last panel shows for the same region of 4 January 2013 the MSSIM curve for the masked areas in the active regions at different wavelengths with the colour coding as in the previous image: 171\,\AA\, in yellow, 193\,\AA\, in blue, 304\,\AA\, in red, and 211\,\AA\, in green.}
 \label{loopfilling}
\end{figure}

\begin{sidewaysfigure}
     \includegraphics[width=0.85\textwidth]{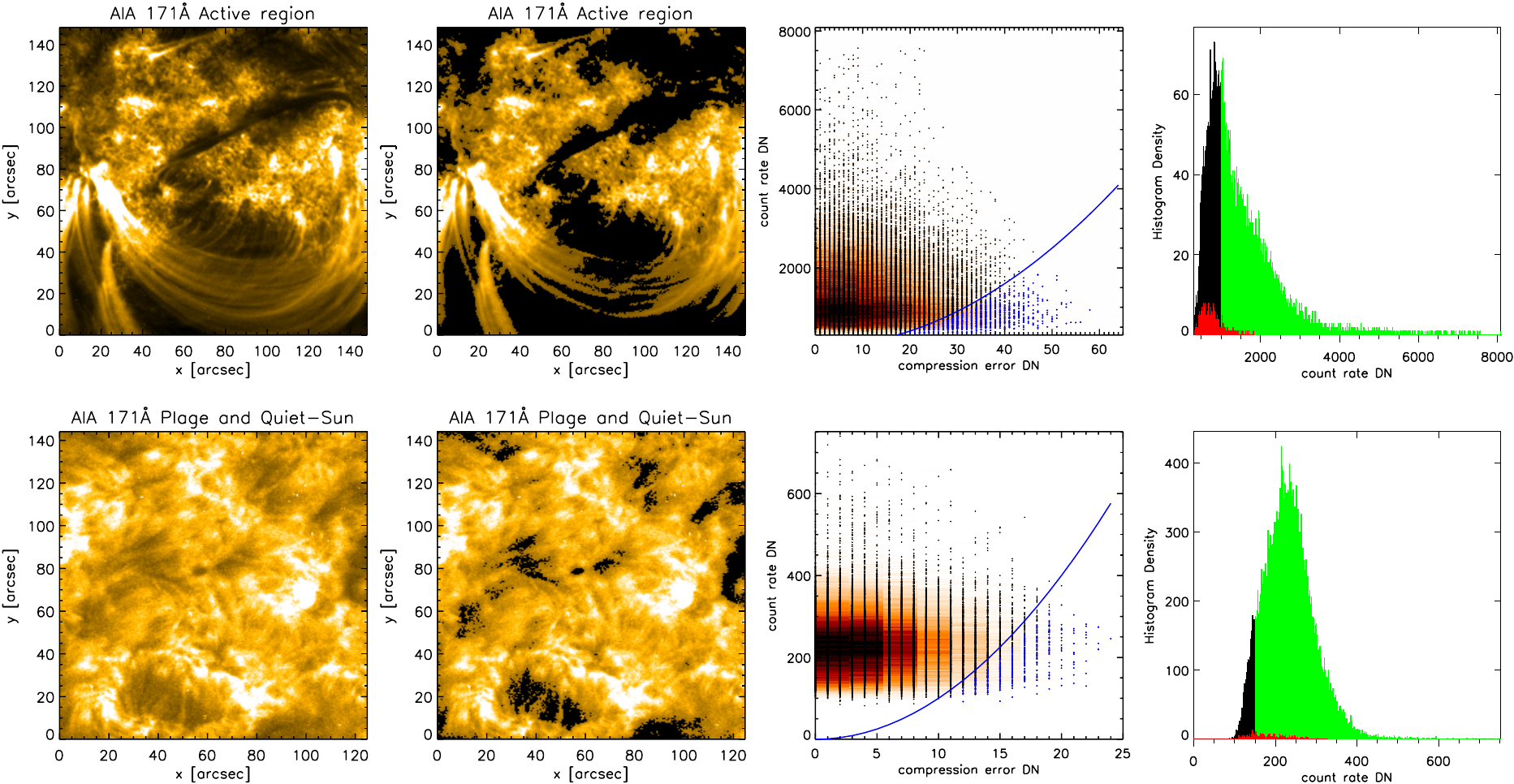}
    \caption{Upper row: The first two panels show an AIA/SDO image of an active region taken in 171\,\AA\, and the same image with image values smaller than a cutoff value of 1000 set to zero. In the third panel we show for a relative compression ratio of three a scatterplot of count rate {\em versus} compression error. 
     In the background, with a red colour table, we show the 2D-histogram calculated for a bin size of three. Locations at which the compression error exceeds the photon noise are indicated with blue scatter plot points. The solid blue line marks the square of the compression error.
The last panel in the row shows the histogram of the count rates in the image for the entire original image in black for pixel intensity values above 1000 in green and for pixels where the photon noise is exceeded by the compression error for the image compressed with a CR$_{rel}$=3 in red. Lower row: Same as in upper row but now for a quiet-Sun and plage region in 171\,\AA\, and with a mask cutoff value of 150.}

   \label{pix_dens}
    \end{sidewaysfigure}

In order to take this into account, we plot in Figure~\ref{loopfilling} the MSSIM curve with respect to the bitrate, now only for the active regions recorded in 171\,\AA. The qualitative shape is the same for all regions -- a slow decrease in the MSSIM value with bitrate, followed by a faster decline with the turning point at different locations for the varying regions. The image to maintain the higher quality even at low bitrates (high compression) is the image taken on 4 January 2013 (dotted curve). This image is almost entirely filled with active-region loops. To demonstrate the MSSIM dependance on such a ``structure-filling'' factor, we plot in an inset of the first panel the MSSIM for a relative compression ratio of three. The more the image is filled with loops, the higher the MSSIM at a given relative compression ratio. We then proceed to mask the images by choosing a cut-off value of 1000\,DN. All pixels below this value are set to zero. In this way, we obtain the location of pixels we believe to contain the structure and intensity variations that would be used when scientifically analysing the image (see Figure~\ref{pix_dens} for an example of such a masked image). We plot in the second panel of  Figure~\ref{loopfilling}  the MSSIM curves again, but when averaging over the SSIM map (obtained by Equation~(\ref{SSIMeq}) using a Gaussian window), we only take the pixels located within the mask into account. The retrieved curves then behave quantitatively more similarly. The vertical lines denote the bitrates at which a MSSIM of 0.95 is produced for each of the regions, and the bitrates range from to $0.4$ to $1.2$.  In the last panel, the MSSIM curves obtained in this way for the region recorded on 4 January 2013 are shown for all four wavelengths with a mask cut-off value of 700 for the 193\,\AA\,  regions,  500 for the 304\,\AA\,  regions, and 150 for the 211\,\AA\,  regions. The intent is to eliminate the dependance of the MSSIM value on the actual amount of scientifically  relevant signal in the image. The vertical lines again denote the bitrates at the MSSIM value of 0.95, which we believe to be a reasonable cutoff for each wavelength as we stay above the rapid decay of the image quality.

\begin{table}
\caption{The table lists the data set (wavelength and date, see also Table~\ref{database}) followed by the relative compression ratio and bitrate at an MSSIM of 0.95 for the active regions and 0.85 for the quiet-Sun regions. For active regions the MSSIM value has been extracted from the curve calculated using only pixels that are within the masked area (see also text for explanation). The fourth column shows the percentage of pixels that have a compression error larger than the photon noise, and for active regions, the last column is the percentage of pixels showing a compression error larger than the photon noise and that additionally are located within the masked area. Values above 15\,\% are shown in bold face.}

\label{crtable}
\begin{tabular}{l c c c c }
 \hline
  \hline
 && Active regions& MSSIM=0.95&\\
 \hline
 Data & CR$_{rel}$& Bitrate  &  
ce  $>$\,noise [\%] & ce  $>$\,noise [\%]    \\ 
&  &  & total image &   masked image  \\ 
AR 171/ 4 Jan. 2012 & 7.0 & 1.1 & 16.8 & 9.1 \\
AR 171/ 4 Jan. 2013 & 9.0 & 0.9 & 29.0 & {  \bf 27.2} \\
AR 171/19 Jan. 2012 & 13.0 & 0.6 & 41.6 &  {  \bf 33.5} \\
AR 171/28 Mar. 2013 & 16.0 & 0.4 & 29.6 & { \bf 22.1} \\
AR 193/ 4 Jan. 2012 & 4.0 & 1.9 & 3.2 & 1.6 \\
AR 193/ 4 Jan. 2013 & 6.0 & 1.4 & 12.0 &10.8\\
AR 193/19 Jan. 2012 & 8.0 & 0.9 & 18.9 & 13.5 \\
AR 193/28 Mar. 2013 & 6.0 & 1.2 & 7.4 & 2.7 \\
AR 304/ 4 Jan. 2012 & 5.0 & 1.1 & 18.8 &  11.4\\
AR 304/ 4 Jan. 2013 & 5.0 & 1.3 & 29.7 & { \bf 26.7}\\
AR 304/19 Jan. 2012 & 5.0 & 1.2 & 19.8 & 12.1\\
AR 304/28 Mar. 2013 & 5.0 & 1.1 & 19.4 &  11.6\\
AR 211/ 4 Jan. 2012 & 4.0 & 1.7 & 4.9 & 1.0 \\
AR 211/ 4 Jan. 2013 & 6.0 & 1.2 & 11.7 & 7.0 \\
AR 211/19 Jan. 2012 & 7.0 & 0.9 & 11.5 & 4.7 \\
AR 211/28 Mar. 2013 & 7.0 & 0.9 & 11.0 & 2.8 \\
 \hline
  \hline
 && Quiet sun& MSSIM=0.85&\\
 \hline
 Data &  CR$_{rel}$& Bitrate  &  
ce  $>$\,noise [\%] &  \\ 
&  &  & total image &  \\ 
QS 171/ 4 Jan. 2012 & 6.0 & 1.1 & 9.9 &  \\
QS 171/ 4 Jan. 2013 & 5.0 & 1.3 & 4.6 &  \\
QS 171/19 Jan. 2012 & 4.0 & 1.4 & 5.9 &  \\
QS 171/28 Mar. 2013 & 5.0 & 1.2 & 5.3 &  \\
QS 193/ 4 Jan. 2012 & 4.0 & 1.5 & 1.5 &  \\
QS 193/ 4 Jan. 2013 & 4.0 & 1.5 & 2.1 &  \\
QS 193/19 Jan. 2012 & 4.0 & 1.4 & 3.3 &  \\
QS 193/28 Mar. 2013 & 4.0 & 1.5 & 1.9 &  \\
QS 304/ 4 Jan. 2012 & 4.0 & 1.2 & 10.8 &  \\
QS 304/ 4 Jan. 2013 & 4.0 & 1.2 &10.7 &  \\
QS 304/19 Jan. 2012 & 3.0 & 1.5 & 6.0 &  \\
QS 304/28 Mar. 2013 & 4.0 & 1.2 & 10.7&  \\
QS 211/ 4 Jan. 2012 & 3.0 & 1.8 & 0.7 &  \\
QS 211/ 4 Jan. 2013 & 3.0 & 1.8 & 1.6 &  \\
QS 211/19 Jan. 2012 & 3.0 & 1.7 & 1.5 &  \\
QS 211/28 Mar. 2013 & 3.0 & 1.7 & 1.1 &  \\
  \hline
    \hline
\end{tabular}
\end{table}

We use the cut-off bitrates found in this way to obtain Table~\ref{crtable}, in which we list relative compression ratios for the EUV images at the different wavelengths by choosing a recommended MSSIM for the masked active regions of 0.95 and an MSSIM for the quiet regions of 0.85. The percentage of pixels exhibiting a compression error larger than the photon noise (for the active regions listed for the entire image as well as for the masked image) gives an additional constraint on the quality. 
The image noise for the AIA level 1 data at different wavelengths is calculated with the {\bf aia\textunderscore bp\textunderscore estimate\textunderscore error.pro} IDL routine provided by the AIA instrument team. The various noise sources (photon noise, read-out noise, and so on) are explained in \cite{2012SoPh..275...41B}, where it is also shown that the photon noise is by far the dominant factor. 

We have marked in Table~\ref{crtable}, for the percentage of pixels with a compression error greater than the photon noise, the values above 15\,\% in bold face. The 171\,\AA\, wavelength shows the highest percentage of pixels with compression errors greater than the photon noise. However, as we demonstrate in the first row of Figure~\ref{pix_dens}, this is acceptable when the only interest is in coronal loops because these are the predominant features of the images and are the last to be affected by compressions. 
In the first row of Figure~\ref{pix_dens} we show just such an active region and the residual image when applying a mask with a cut-off value of 1000. From the compression error {\em versus} count rate plot in the third panel, it is clear that at a relative compression ratio of three the bulk of the pixels still exhibits higher values than the compression error. The histograms of the image in the last panel clearly show that the pixels from high-intensity high-structure regions (including location of loops) at this compression rate are predominantly not yet members of the group of pixels with a compression error greater than the photon noise. 
In comparison, in a quiet-Sun region at the same wavelength, as shown in the second row of Figure~\ref{pix_dens}, the structures fill almost the entire image. They have low intensity values, and the compression error does not seem to scale with the intensity values. While the bulk of the image values is about a fifth smaller than in the active region, the compression error only reduces its range by a factor of two. 

\subsection{Scientific Test Cases }\label{sec:resultsclo} 

For scientific data analysis of EUV solar images, brightness variations in the image and displacements or movement of structures (such as coronal loops) in a time series become important in deriving physical parameters. 
It therefore becomes critical to study how the compression scheme handles sharp edges that can significantly alter the scientific result. The typical ringing artefact encountered during JPEG2000 compression is exactly this weak point in the compression scheme because it artificially imposes an oscillation at the sharp edge by the so-called over- and undershooting when dealing with spatially bandwidth limited data.

We investigate the influence of the relative compression ratios on the derived pixel intensity displacement in coronal loop oscillations and find the consequences for the retrieved oscillation parameters.

\subsubsection{Coronal Loop Oscillation: Off-limb}\label{sec:resultsclo_1}

 \begin{figure*}
\includegraphics[width=\textwidth]{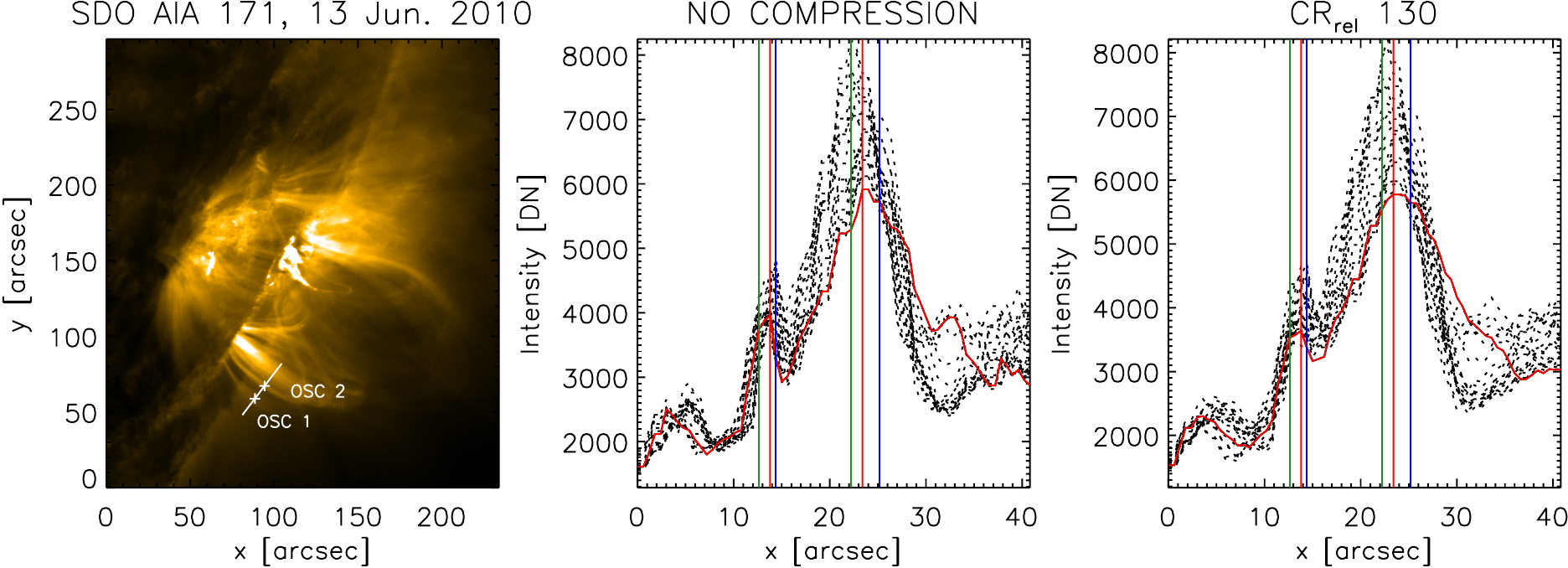} 

  \caption{Left to right: One image of the time series of the coronal-loop oscillation observed on 13 June 2010  with AIA/SDO in 171\,\AA\, between 05:30 UT and 05:50 UT.
  The white line indicates the cut shown in the following panels. The two  analysed oscillations are indicated with the white cross as OSC 1 (isolated loop) and OSC 2 (loop in loop bundle). The second panel shows the profile along the white line of the first panel for the uncompressed time series. The profile from left to right first shows OSC1 and then OSC2. The red solid line is the starting profile, and the dashed lines show the ensuing time steps. The red vertical lines mark the loop positions at the first time step, and the blue and green vertical lines show the maximal transverse loop displacement during the time series for both sides, respectively.  The last panel is similar to panel 2 but with the time series images compressed with a relative compression ratio of 130.}

 \label{corosci}
\end{figure*}
 \begin{figure*}
\includegraphics[width=\textwidth]{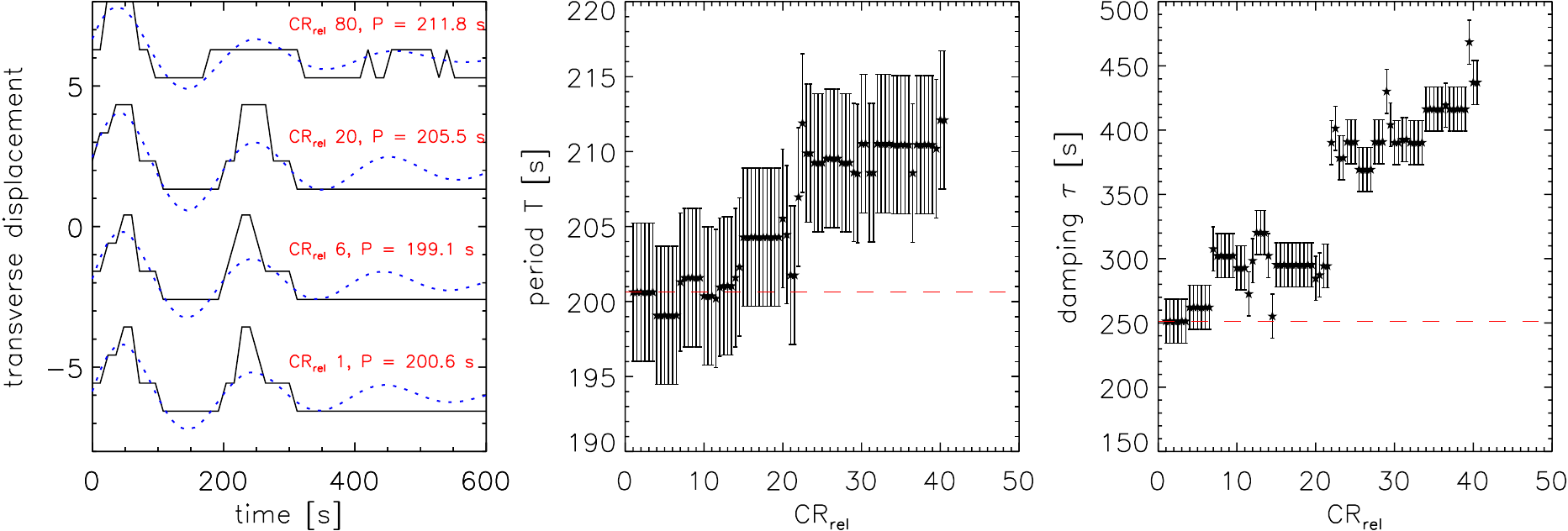} 

  \caption{ Left to right: The first panel shows the retrieved oscillation for different relative compression ratios shifted in the y-axis by an arbitrary amount for easy viewing. The blue dashed lines are the fitted functions according to Equation~(\ref{oscieq}). The second and third panels show the retrieved oscillation period $P$ and the damping $\tau$ from the fit for increasing relative compression ratios (black stars). The error bars are at  $2.3\%$ of the original $P$ and at $6.8\%$ of the original $\tau$ in accordance with~\cite{2012A&A...537A..49W}. 
  } 
 \label{fitosci}
\end{figure*}

We have selected Active Region NOAA 11079, which was at the solar limb on 13 June 2010 and underwent an M1.0 class flare in the extracted time series. The consequent coronal loop oscillations have previously been analysed by \cite{2012A&A...537A..49W} to derive the physical parameters associated with the loops. The cadence of the 171\,\AA\,AIA/SDO data chosen for analysis was 12\,s, and the time series was rotation corrected.

In Figure~\ref{corosci}, we show the region undergoing the oscillations and plot the time development of the loop profile along the cut marked in the left panel for the losslessly compressed images and the highly compressed images at CR$_{rel}$ 130. It is remarkable that even at a compression rate of 130, the loops can be identified, still showing a similar range of transversal displacement, and the oscillation would therefore still be visible in a browsing tool employing a high relative compression ratio for data-mining purposes. We should note, however, that in the off-limb oscillation the loops are better defined with respect to the background, and this situation is different when searching for oscillations on disk.

To study the effect of compression on the retrieval of physical parameters, we also studied the oscillation in detail by
\begin{enumerate}[label={(\arabic*)},itemsep=0.0cm]
\item taking a cut through the chosen loop, which would be the loop undergoing OSC1 in our case (see Figure~\ref{corosci}),
\item storing for each profile the loop position in time that was found by automatic maximum finding (given a certain range of the profile) and thus retrieving the transverse displacement oscillation, and
\item  fitting the found oscillation  with a damped cosine function (as was done in \cite{2012A&A...537A..49W})
          \begin{equation}
          \xi_{(t)}=\xi_{0}exp \left(  \frac{-  \left( t-t_{0} \right)}{\tau}\right) \cos \left( \frac{2\pi}{P}  \left( t-t_{0}  \right)-\phi \right),
         \label{oscieq}
          \end{equation}
           using the {\bf {mpfit$.$pro}} IDL routine by \cite{2009ASPC..411..251M}, where $\xi$ is the the oscillation amplitude, $P$ the oscillation period, $\tau$ the damping of the oscillation, and $\phi$ the phase.
\end{enumerate}
   We performed these steps for a range of relative compression ratios from 1 to 40.\\

In Figure~\ref{fitosci} we plot the fitted oscillation periods, $P$, and damping parameters, $\tau$, for increasing relative compression ratio CR$_{rel}$.  The error bars we apply are the same percentage in error as was determined by the analysis of \cite{2012A&A...537A..49W} for these parameters. The oscillation periods remain within the error bars of the losslessly compressed data up to a CR$_{rel}$ of 19, but the damping parameters have increased to higher values outside the error bars. 

The developments of the oscillation period and the damping are not random, but show a trend to higher values with increasing compression. As the images become blurred and the intensity differences between pixels is reduced, the location of the loop (the maximum in the loop profile) can be attributed to a different pixel.  As we show in the first panel of Figure~\ref{fitosci} for the oscillation at CR$_{rel}$ 80, additional transverse displacement can be registered even after the oscillation has stopped (from 400\,s onwards). The fitting routine tries to include these points and the amplitude reduction between 180 and 220\,s into the fit, which ``stretches" out the damped cosine function, leading to a higher $P$ and $\tau$.

\subsubsection{Coronal Loop Oscillation: On-disk}\label{sec:resultsclo_2}

 \begin{figure*}
\includegraphics[width=\textwidth]{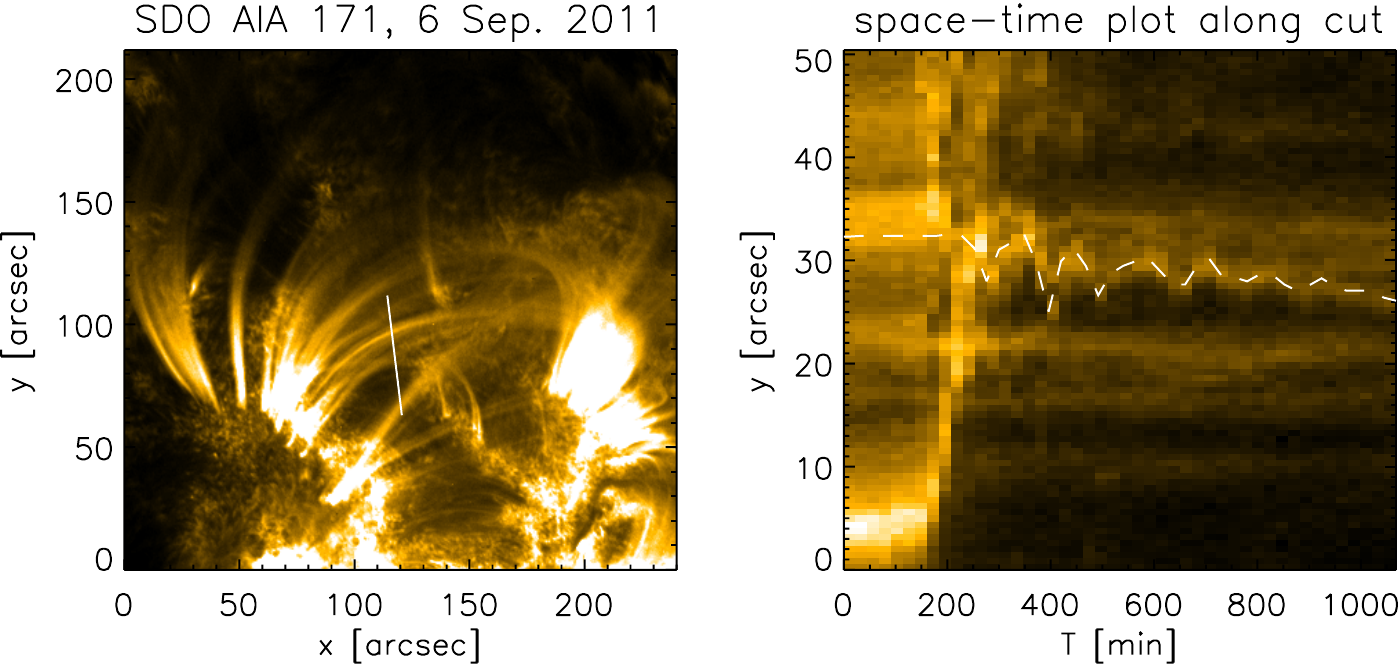} 

  \caption{On the left side we show one image of the time series we analysed of the coronal loop oscillation observed on 6 September 2011.  
  The white line indicates the cut, of which we display a space-time plot on the right side. The white dashed line on the right side shows the transverse displacement found by eye that is analysed further in Figure~\ref{fitosci_disk}.} 
 \label{corosci_disk}
\end{figure*}

\begin{figure*}
\includegraphics[width=\textwidth]{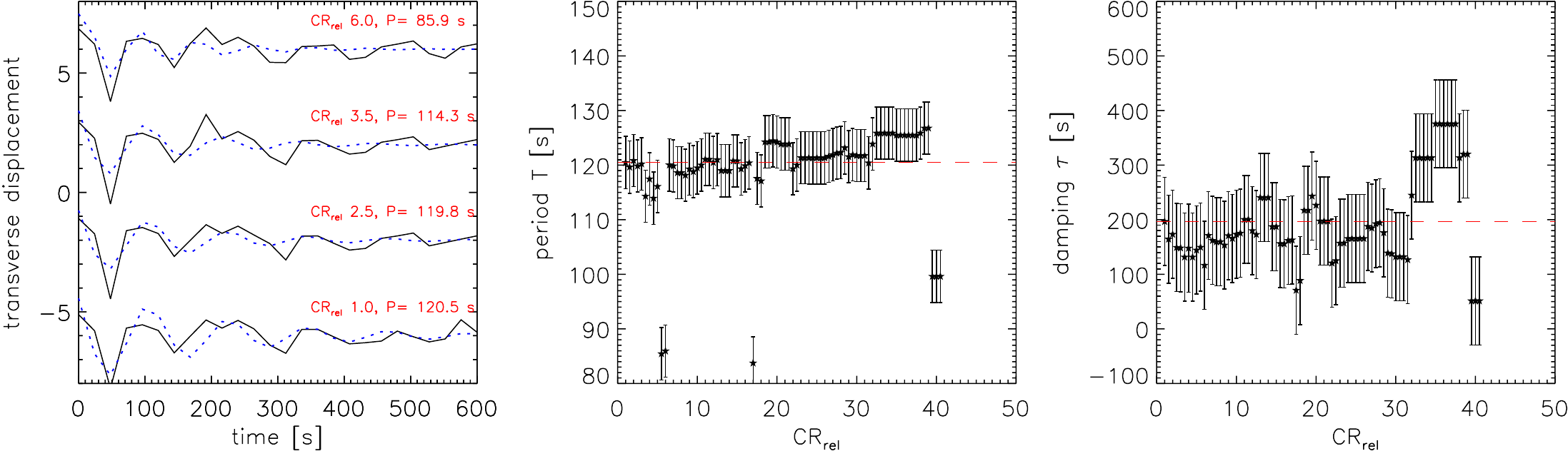} 

  \caption{ Same as in Figure~\ref{fitosci} but now for the on-disk coronal loop oscillation fits and for different relative compression ratio.}
 \label{fitosci_disk}
\end{figure*}

For the on-disk loop oscillation we choose an event seen in AIA/SDO data that was described and analysed by \cite{2015ApJ...804L..19J}. The studied Active Region NOAA\,1283 underwent an X2.1 class flare that peaked at  22:20 UT and lifted a group of coronal loops in the process, with an oscillation being triggered in a group of lower lying loops underneath. We follow their time-series analysis and study the coronal loop oscillations in the lower lying loops. The left side of Figure~\ref{corosci_disk} shows an image taken at 22:00 UT in the 171\,\AA\, line just before the onset of the flare, with a white line indicating our cut through the oscillating loops. On the right the space-time plot shows the visible oscillation of a loop around 32\,arcsec. We have traced the loop position by showing the transverse displacements with a dashed white line. In Figure~\ref{fitosci_disk} we display, as in Figure~\ref{fitosci}, the resulting fits and fit parameter for the progressively more compressed images. The technique is identical to that of Section~\ref{sec:resultsclo_1}, with the exception that the loop positions were found by eye in the uncompressed time series, and subsequently, in the automated routines for the compressed time series, the maximum was not searched for along the entire cut, but in a small range of five pixels around the loop position found in the non-compressed time series. We obtain the errors from the fitting routine by providing an estimate of our error in determining the loops position (two pixels).
We find in the original time series of the uncompressed images a period of $P \sim 120\,s$ and $\tau \sim 200\,s $, which is, for the period, comparable to the fit parameters found by \cite{2015ApJ...804L..19J}; these were about  $P = 120\,s$ and $\tau =300\,s$ for the different studied oscillations.
The period can in general (as was the case for the off-limb oscillations) be retrieved even at high compression, but the fitting routine in this case failed twice at a relative compression ratio of $\sim$ 5.5, which illustrates how fragile the process is to slight changes in intensity values. The damping parameter shows a large scatter with values differing by 50\,s and more between relative compression ratio increases of less than 0.5. There is still an overall trend to larger damping parameters with higher compression, similar to what is observed in the off-limb loop oscillation. It is again surprising, however, that the transverse displacement is still detectable with a period within the error bars up to a compression of around 30.
However, already at a compression of two, the damping parameter is not retrieved within the errors.

\section{Summary} \label{sec:concl}

We have compressed solar EUV images of the AIA/SDO instrument with the JPEG2000 algorithm using varying compression rates. In Section~\ref{compeff}, we demonstrated the importance of the relevant spatial scales and intensity gradient in the images. The smoothing or blurring effect of the compression and the introduced quantisation noise affect the various regions differently:  while coronal loops are still identifiable at high compression rates (around CR$_{rel}$ = 50), the plage and mossy areas already show a loss of structure at relative compression ratios of around CR$_{rel}$=20.
In the next step, we computed the MSSIM and PSNR as quality metrics and compared their behaviours.

In Section~\ref{wvl_mssim}, we compared the MSSIM curves for different wavelengths and found a systematic behaviour. Given the same region on the Sun, 171\,\AA\, images can be compressed with higher compression rates than, for example, images in 304\,\AA\, while maintaining a higher quality (less structure loss). The difference in the original images is that the 171\,\AA\, images show higher intensity values that predominantly trace the coronal loops.
Another aspect is the comparison of the compression noise to the photon noise inherent to the images. We found that within an image the low intensity values are first affected in raising their compression error above the photon noise. For active region images in 171\,\AA\,, this means that pixels belonging to loops (high intensity values) are only affected at high compression rates.

We find that there is not a single, general relative compression ratio recommendation for images at a certain wavelength. Depending on the structure scale of interest, the intensity gradients of the structure and the nature of solar events searched for, different compression ratios are feasible. Nevertheless, by introducing a simple masking of the images (for the active regions), we obtain in Table~\ref{crtable}, for a chosen MSSIM recommended relative compression ratios for different regions and wavelengths.
However, we point out that these numbers are only relevant to the spatial resolution and sensitivity of the chosen specific instrument because with increasing or decreasing spatial resolution, for instance, the fine structure will change. For a pixel scale of 0.6\,arcsec, for example, with a scientific interest in retrieving physical parameters of coronal loop oscillations in 171\,\AA\, from off-limb data, a relative compression ratio of around six is achievable, which ensures remaining at around 20$\%$ of the typical photon noise, while higher relative compression ratios are possible for browsing (range CR$_{rel}$ \rm{10\,--\,12}). The results obtained here can therefore not be directly applied to {\em Solar Orbiter} {\em EUI/High resolution imager} data, for example,  which will have a resolution higher by about five times than the resolution of the AIA/SDO images. 

The coronal loop oscillations on-disk studied in Section~\ref{sec:resultsclo_2} also showed that, whilst the transverse displacement is still observed and detected at relative compression ratios as high as CR$_{rel}$=30, the physical parameters cannot be reliably determined in that case from the compressed data. This  of course depends on the amplitude of the oscillation (transverse displacement of the loop) and the background intensity and on whether there are any overlying loops in the line-of-sight.  

The aim of this study was to assess the possibility of increasing data storage and transfer efficiency for large solar databases using a compression algorithm, while also outlining approaches that might be relevant for on-board data compression for telemetry-constrained space missions like {\em Solar Orbiter}. This is a complex issue involving a wide range of image- and signal-processing topics, and our findings underline the fact that the intended usage of the data to be compressed plays an important role in deciding on the compression limits. 
With the flood of high-resolution data to be expected from the new generation of large solar telescopes, such as the {\em Daniel K. Inouye Solar Telescope}, these type of studies is essential. The growing number of sessions and meetings devoted to the handling of ``big data" in solar physics shows a rising awareness in the solar community, and we hope this contribution will encourage further studies in this field.

                    


\section*{Acknowledgements}

AIA/SDO data are provided by the Joint Science Operations Center--Science Data Processing. This project has received funding from the Science and Technology Facilities Council (UK) and the European Research Council (ERC) under the European Union's Horizon 2020 research and innovation programme (grant agreement No 647214).
I.~De~Moortel and C.~E.~Fischer thank the ESTEC/ESA visitorship program for support.
We have made use of the Coyote IDL library at www.idlcoyote.com.\\

                    


\clearpage

\appendix
\begin{longtable}[h!]{l c c c c   }

\caption[AIA/SDO database. Pixel coordinates are given for the  4096 x 4096 pixel images.]{AIA/SDO database. Pixel coordinates are given for the  4096 x 4096 pixel images.} \label{database} \\
&Time &wavelength&target\\
 &(UTC)&[\AA]& (x1,y1,x2,y2) [pixel]\\
  \hline
  \multicolumn{4}{l} {4 Jan. 2012}\\
 \hline
  
&&&\\
  Example in 171\,\AA:&&&\\
\multirow{4}{*}{\vspace{0.5cm} \includegraphics[width=3cm]{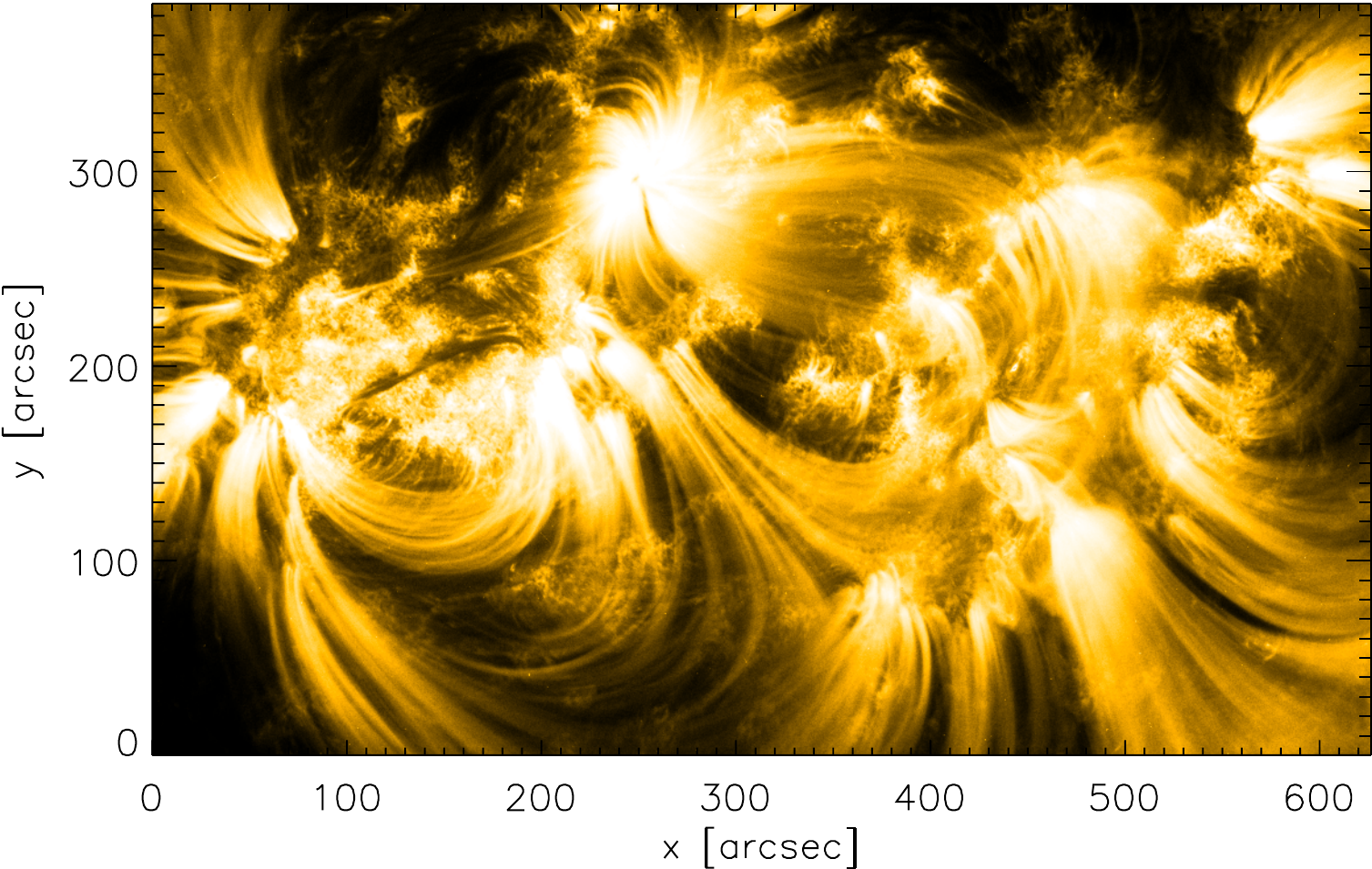}}&12:00:01&171&Active Region\\
&12:00:04&193& (2028,1124,3072,1768)\\
&12:00:09&304&\\
&12:00:02&211&\\
&&&\\

\hline
&&&\\
  Example in 171\,\AA:&&&\\
\multirow{4}{*}{\includegraphics[width=2.2cm]{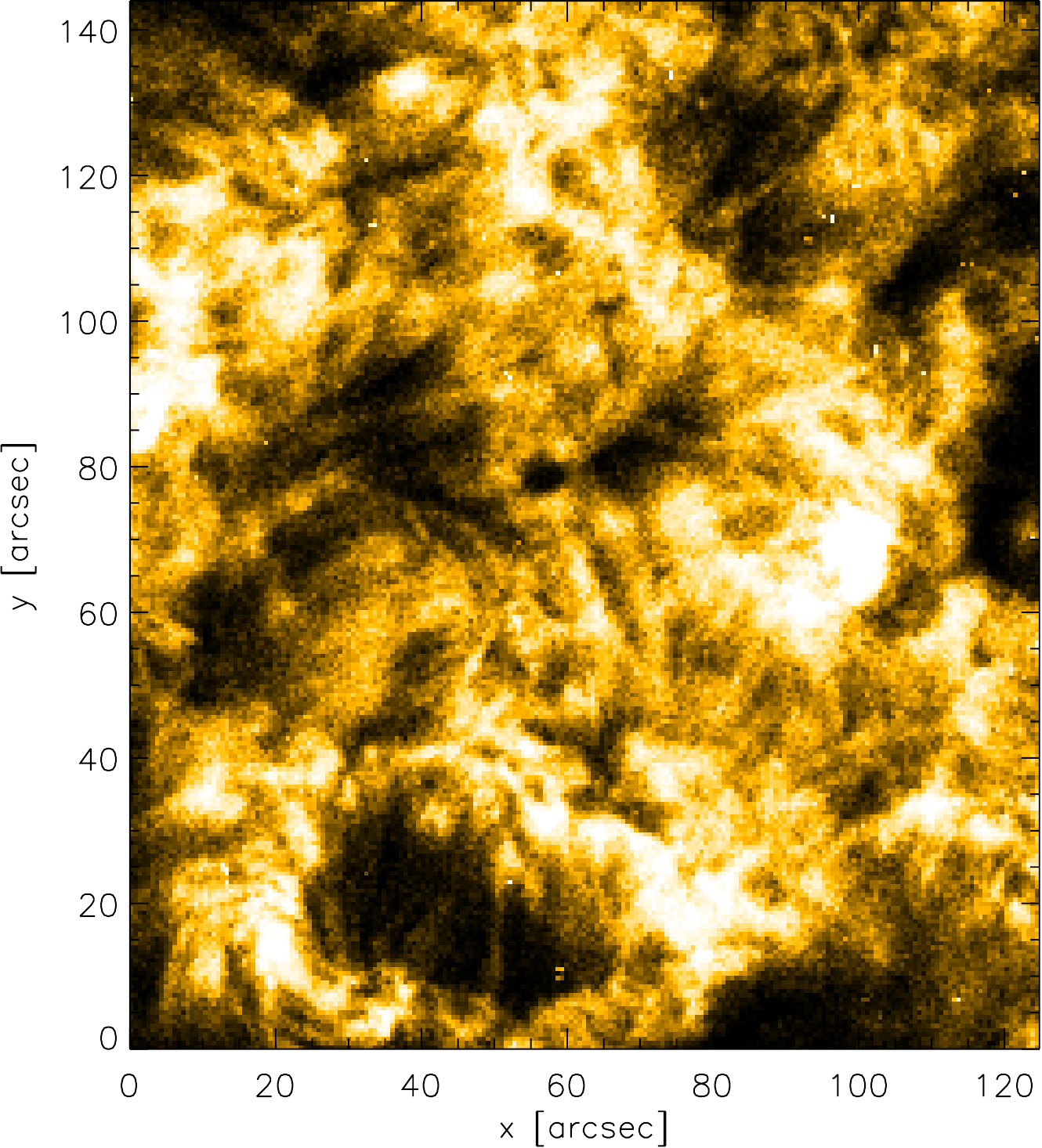}}&12:00:01&171&Quiet Sun \\
&12:00:04&193&(1372,1420,1580,1660)\\
&12:00:09&304&\\
&12:00:02&211&\\
&&&\\
&&&\\
 \hline
\multicolumn{4}{l} {19 Jan. 2012}\\
 \hline
 &&&\\
  Example in 211\,\AA:&&&\\
\multirow{4}{*}{\includegraphics[width=3cm]{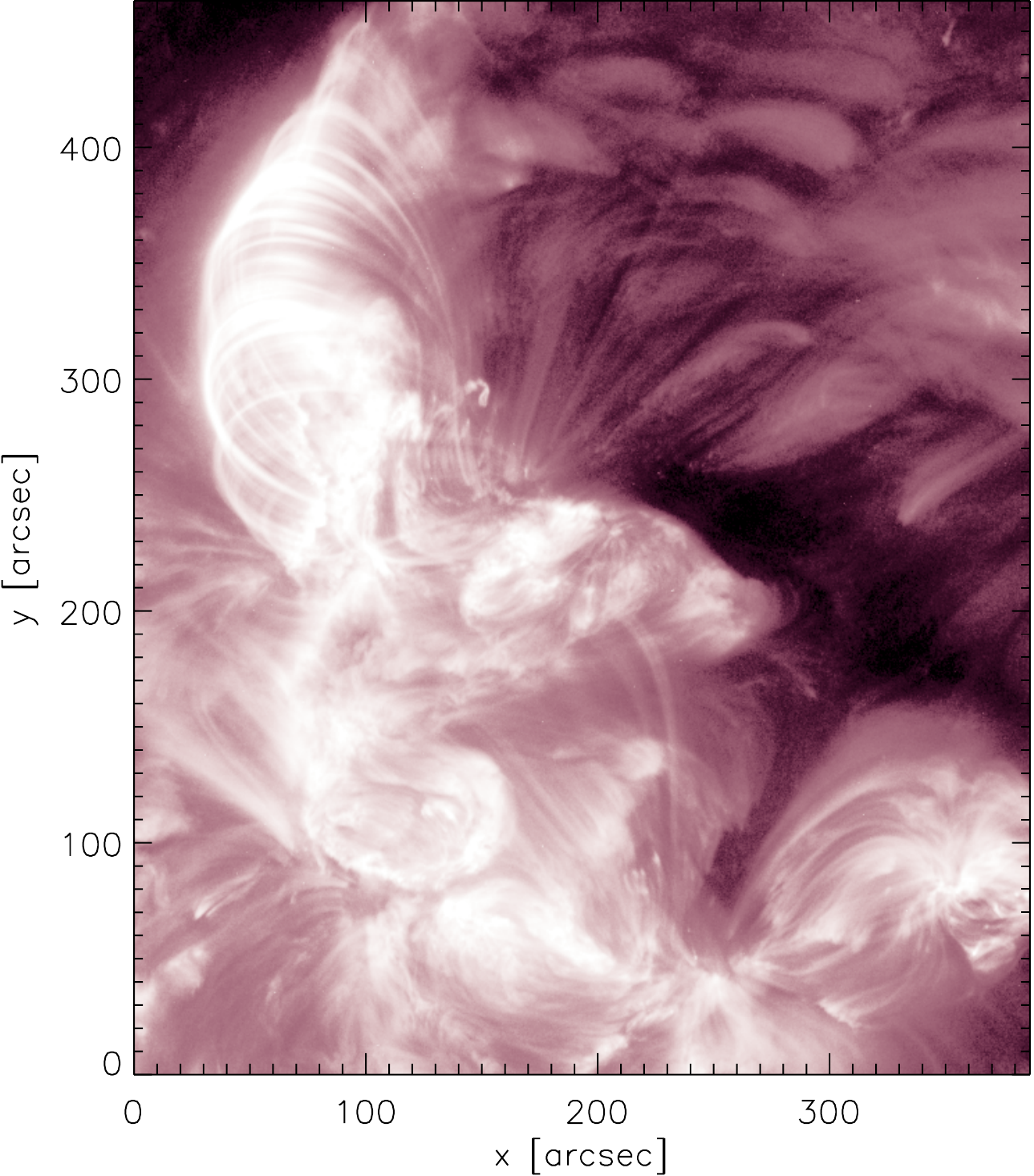}}&21:30:01&171&Active Region\\
&21:30:08&193& (1360,2532,2004,3304)\\
&21:30:09&304&\\
&21:30:02&211&\\
&&&\\
&&&\\
&&&\\
&&&\\
&&&\\
\hline
&&&\\
Example in 211\,\AA:&&&\\
\multirow{4}{*}{\includegraphics[width=3cm]{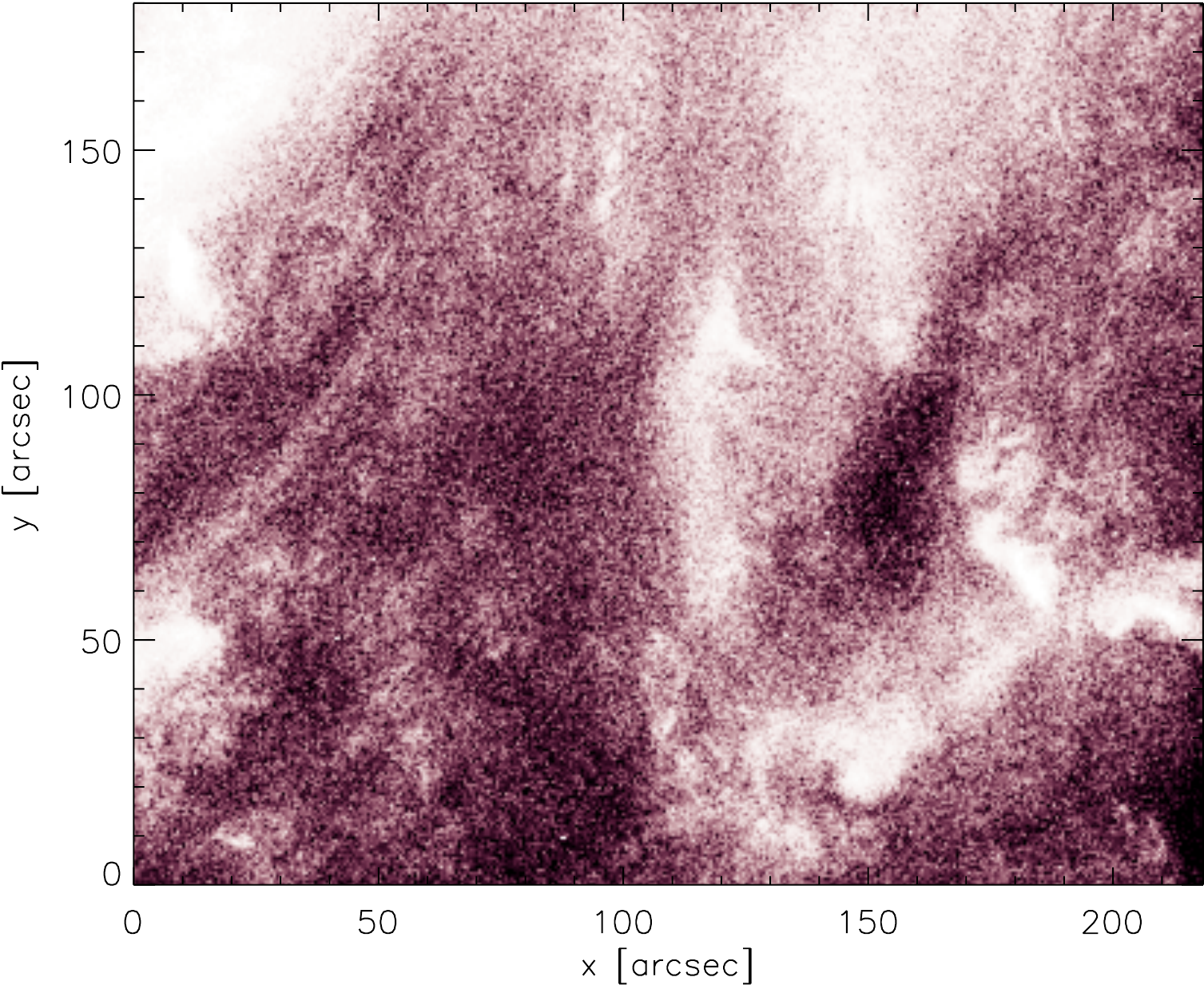}}&21:30:01&171&Quiet Sun\\
&21:30:08&193& (1360,1900,1724,2200)\\
&21:30:09&304&\\
&21:30:02&211&\\
&&&\\
&&&\\
&&&\\
  \hline
&&&\\
&&&\\
&&&\\
&&&\\
&Time &wavelength&target\\
 &(UTC)&[\AA]& (x1,y1,x2,y2) [pixel]\\
  \hline
 \hline
\multicolumn{4}{l} {4 Jan. 2013}\\
 \hline
  &&&\\
  Example in 193\,\AA:&&&\\
\multirow{4}{*}{\includegraphics[width=3cm]{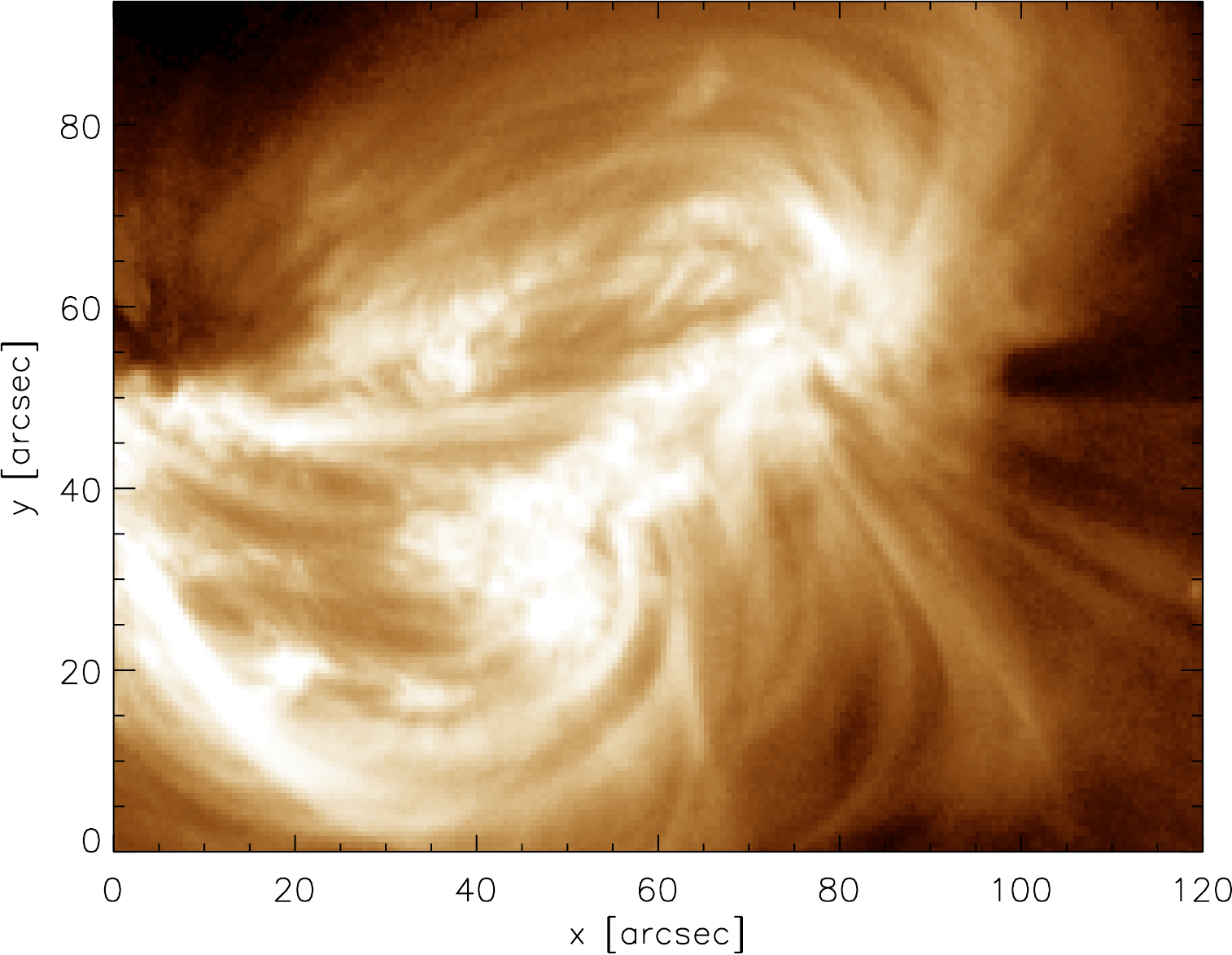}}&14:00:00&171&Active Region\\
&14:00:07&193&( 2292,1688,2492,1844)\\
&14:00:08&304&\\
&14:00:01&211&\\
&&&\\
&&&\\
&&&\\
\hline
&&&\\
  Example in 193\,\AA:&&&\\
\multirow{4}{*}{\includegraphics[width=3cm]{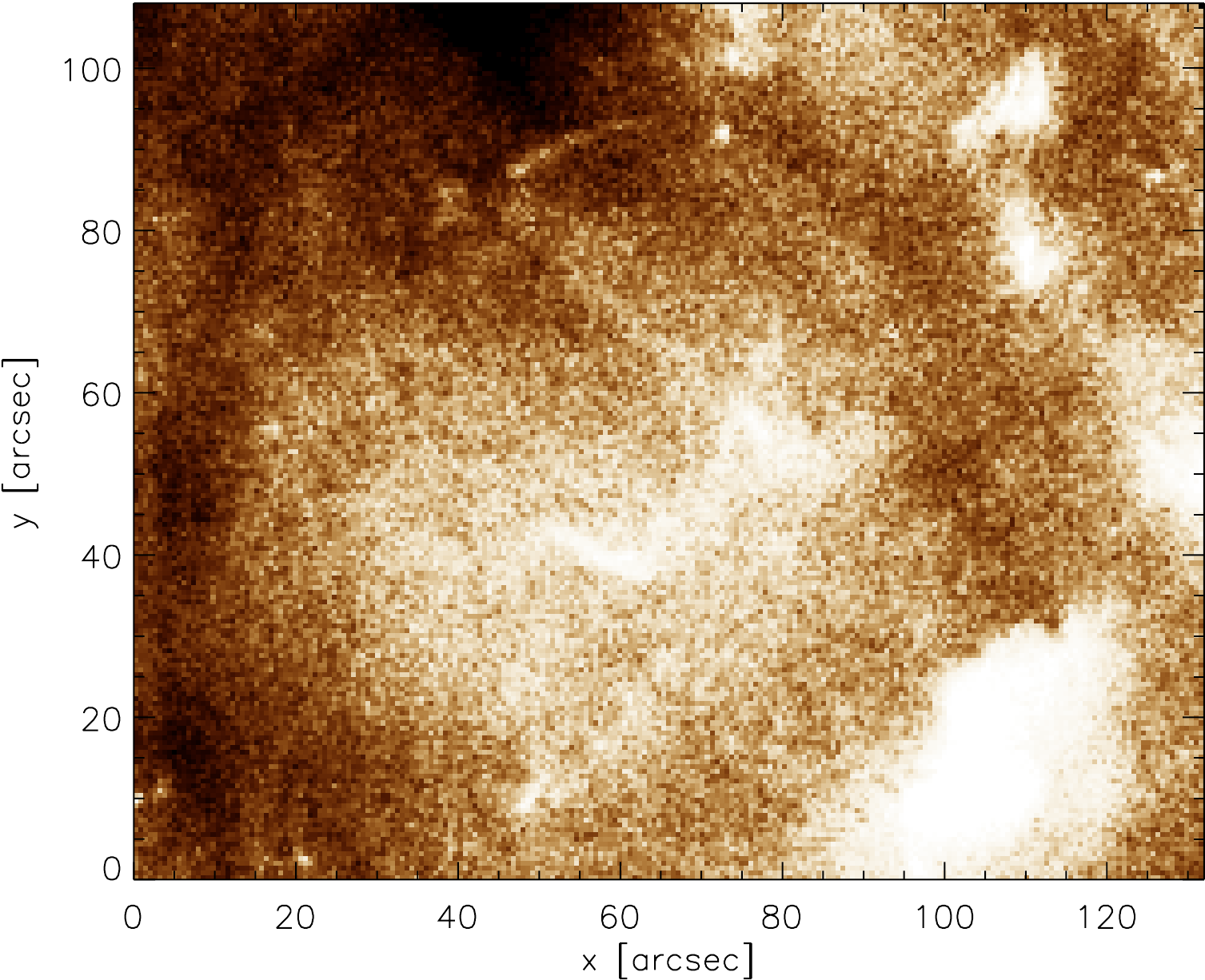}}&14:00:00&171&Quiet Sun\\
&14:00:07&193& (2332,936,2552,1116)\\
&14:00:08&304&\\
&14:00:01&211&\\
&&&\\
&&&\\
&&&\\
 \hline
\multicolumn{4}{l} {28 Mar. 2013}\\
 \hline
   &&&\\
  Example in 304\,\AA:&&&\\
\multirow{4}{*}{\includegraphics[width=3cm]{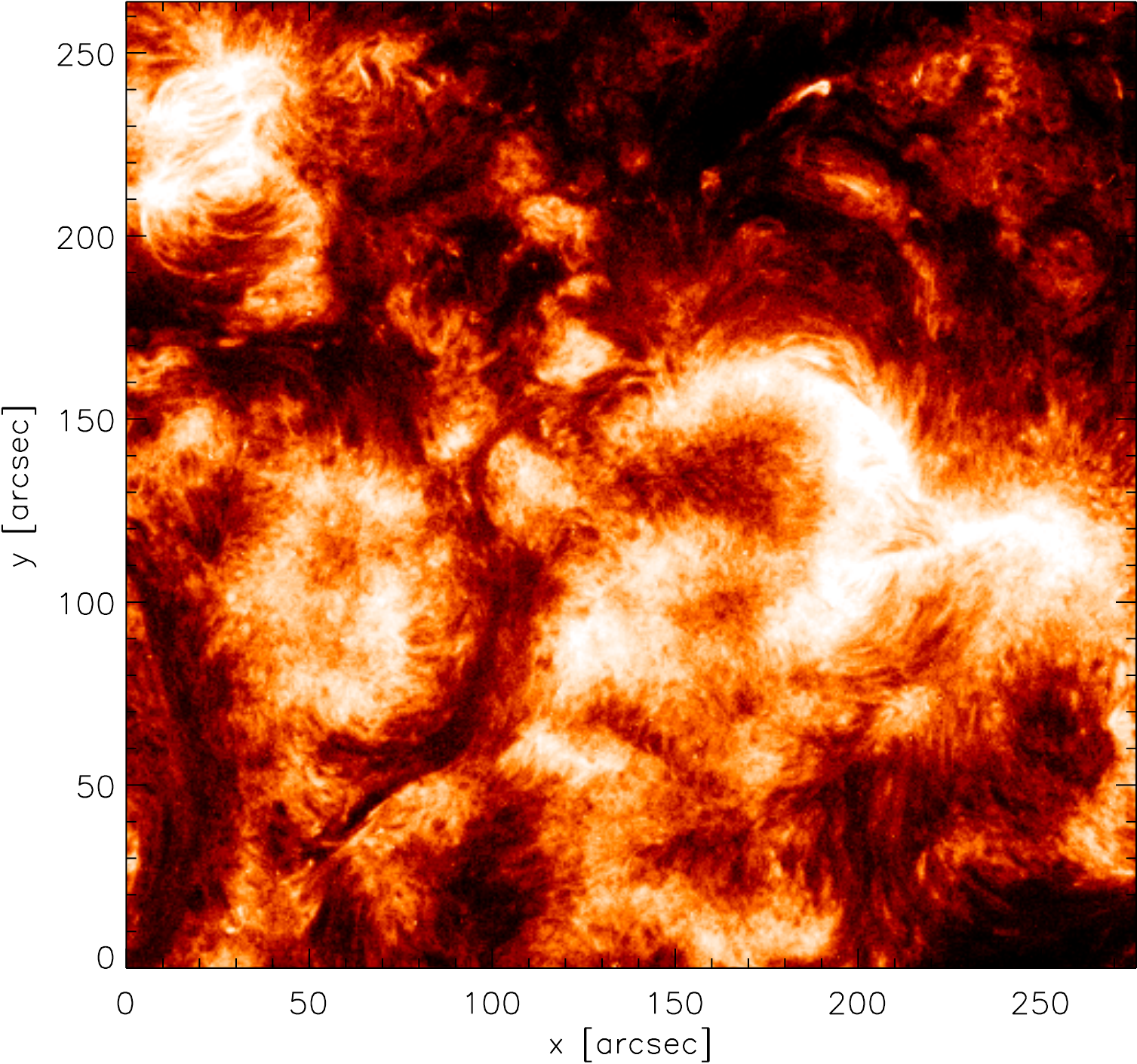}}&12:00:00&171&Active Region\\
&12:00:07&193& (2100,1664,2560,2104)\\
&12:00:08&304&\\
&12:00:01&211&\\
&&&\\
&&&\\
&&&\\
\hline
&&&\\
  Example in 304\,\AA:&&&\\
\multirow{4}{*}{\includegraphics[width=3cm]{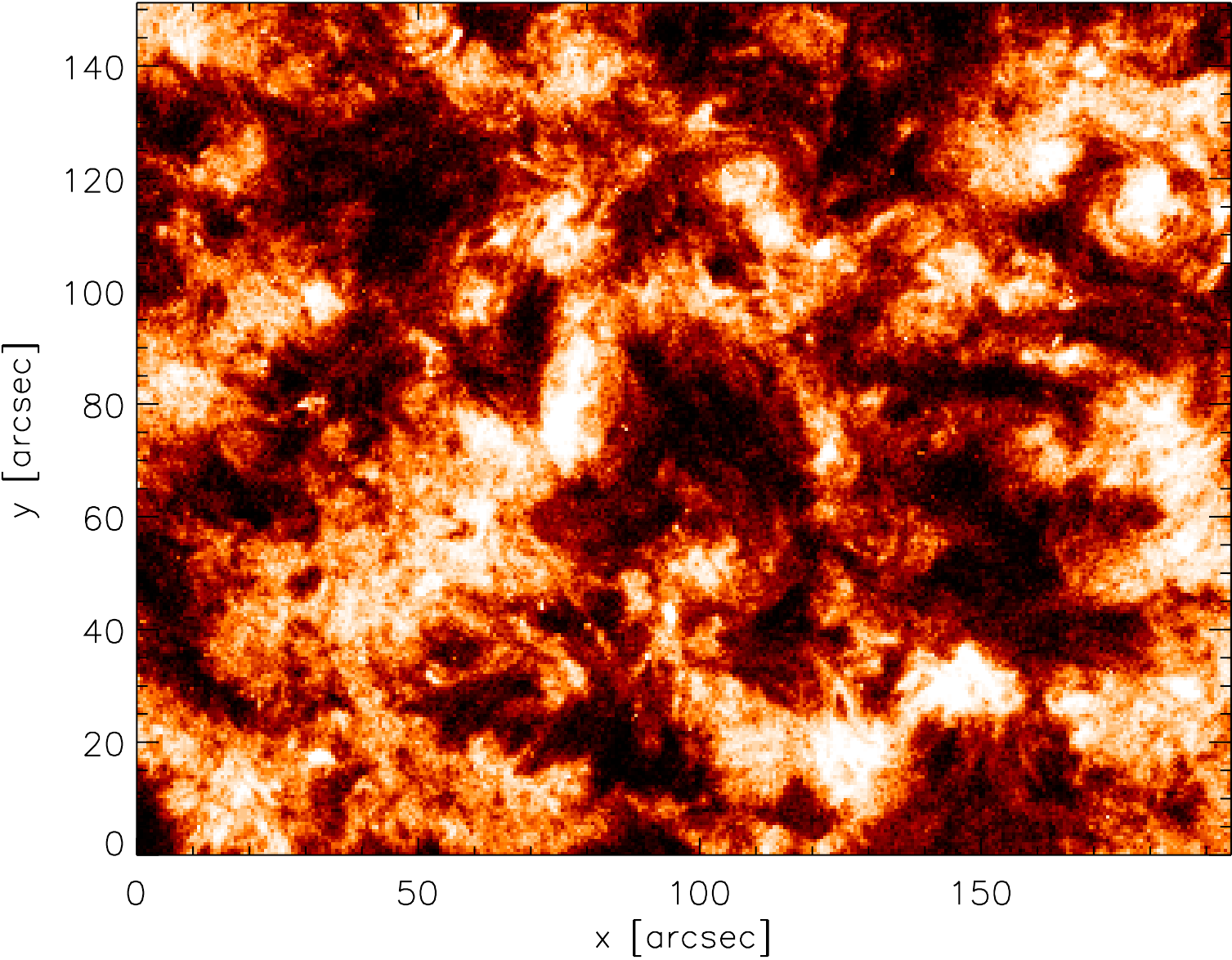}}&12:00:00&171&Quiet Sun\\
&12:00:07&193& (1476,1264,1800,1516)\\
&12:00:08&304&\\
&12:00:01&211&\\
&&&\\
&&&\\
&&&\\
\hline
\end{longtable}

                    


\clearpage
\parskip=-1ex
\bibliographystyle{spr-mp-sola}   
\bibliography{cfischer_bibfile}

\end{document}